\documentclass[
  journal=pasa,
  manuscript=research-paper,
  year=2025,
  volume=XX,
]{cup-journal}

\usepackage{amsmath}
\usepackage[nopatch]{microtype}
\usepackage{booktabs}
\usepackage{graphicx}
\usepackage{amsmath}
\usepackage{amssymb}
\usepackage{gensymb}
\usepackage{lineno}
\usepackage{float}
\usepackage{multirow}
\usepackage{hyperref}
\hypersetup{colorlinks,citecolor=blue,linkcolor=blue,urlcolor=blue}

\newcommand\gray{$\gamma$-ray}
\newcommand\Edot{$\dot{E}$}
\newcommand\Pdot{$\dot{P}$}

\newcommand\LumUnit{erg s$^{-1}$}
\newcommand\Msun{M$_{\odot}$}

\title{Binary population synthesis of the Galactic canonical pulsar population}

\author{Yuzhe Song}
\affiliation{Centre for Astrophysics and Supercomputing, Swinburne University of Technology, Hawthorn, VIC 3122, Australia}
\alsoaffiliation{OzGrav, ARC Centre for Excellence of Gravitational Wave Discovery, Hawthorn, VIC 3122, Australia}
\email[Y. Song]{yuzhesong@swin.edu.au}

\author{Simon Stevenson}
\affiliation{Centre for Astrophysics and Supercomputing, Swinburne University of Technology, Hawthorn, VIC 3122, Australia}
\alsoaffiliation{OzGrav, ARC Centre for Excellence of Gravitational Wave Discovery, Hawthorn, VIC 3122, Australia}

\author{Debatri Chattopadhyay}
\affiliation{Gravity Exploration Institute, School of Physics and Astronomy, Cardiff University, Cardiff, CF24 3AA, United Kingdom}
\alsoaffiliation{Center for Interdisciplinary Exploration and Research in Astrophysics (CIERA) and Department of Physics \& Astronomy, Northwestern University, 1800 Sherman Ave, Evanston, IL 60201, USA}

\author{Joshua Tan}
\affiliation{Department of Natural Sciences, LaGuardia Community College, City University of New York, 31-10 Thomson Ave, Long Island City, NY 11101, USA}
\alsoaffiliation{Department of Astrophysics, American Museum of Natural History, Central Park West at 79th Street, New York, NY 10024, USA}

\author{Timothy A. D. Paglione}
\affiliation{Department of Earth \& Physical Sciences, York College, City University of New York, 94-20 Guy R. Brewer Blvd., Jamaica, NY 11451, USA}
\alsoaffiliation{Department of Astrophysics, American Museum of Natural History, Central Park West at 79th Street, New York, NY 10024, USA}
\alsoaffiliation{Department of Physics, The Graduate Center, City University of New York, 365 Fifth Ave., New York, NY 10016, USA}

\keywords{Pulsars: general, stars: massive, binaries: general, gamma-rays: general } 

\begin{document}

\begin{abstract}
Pulsars are rapidly rotating neutron stars that emit radiation across the electromagnetic spectrum, from radio to \gray s.
We use the rapid binary population synthesis suite COMPAS to model the Galactic population of canonical pulsars.
We account for both radio and \gray\ selection effects, as well as the motion of pulsars in the Galactic potential due to natal kicks.
We compare our models to the catalogues of pulsars detected in the radio, and those detected in \gray s by \textit{Fermi}, and find broad agreement with both populations.
We reproduce the observed ratio of radio-loud to radio-quiet \gray\ pulsars.
We further examine the possibility of low spin-down luminosity (\Edot\ ) pulsars emitting weak, unpulsed \gray\ emission and attempt to match this with results from a recent \gray\ stacking survey of these pulsars.
We confirm the correlation between the latitude of a pulsar and its $\dot{E}$ arises due to natal kicks imparted to pulsars at birth, assuming that all pulsars are born in the Galactic disk. 

\end{abstract}


\section{Introduction}
\label{sec:intro}

Pulsars are rotating neutron stars most commonly observed by their periodic radio emission. Around 3500 pulsars are known to date \citep[][]{Manchester:2004bp}. Modern radio facilities such as MeerKAT, ASKAP and FAST are uncovering new pulsars \citep[e.g.,][]{Han:2021RAA,Padmanabh:2023arXiv}, whilst modern search techniques are still discovering pulsars in decades-old archival surveys \citep[e.g.,][]{Sengar:2023MNRAS,Sengar:2025MNRAS}. The ever-growing pulsar population provides an invaluable insight into the final remnants of massive stellar evolution, whilst relativistic binary pulsars allow for exquisitely precise tests of theories of gravity \citep[e.g.,][]{Kramer:2021jcw}.

Whilst most pulsars are discovered and studied through radio surveys, some are also observed at other wavelengths, including optical \citep[][]{Cocke:1969Natur,Nather:1969Natur}, X-rays \citep[][]{Bradt:1969Natur,Fritz:1969Sci,Oegelman:1993Natur} and \gray s \citep[][]{Kniffen:1974Natur,Tumer:1984Natur,Thompson:1992Natur}.
A growing number of pulsars are being detected in \gray s by the Large Area Telescope (LAT) onboard the {\it Fermi} space telescope \citep[][]{Fermi-LAT:2023zzt}, including some pulsars which were first detected in \gray s and then later detected as radio pulsars, a situation exemplified by the Geminga pulsar \citep[][]{Halpern:1992Natur,Bertsch:1992Natur,Malofeev:1997Natur}.

The spindown luminosity, \Edot\ , is a useful parameter describing a pulsar, and is defined as 
\begin{equation}
    \dot{E} = 4 \pi^{2} I \dot{P} P^{-3},
    \label{eq:Edot}
\end{equation}
where $P$ is the spin period of the pulsar, $\dot{P}$ is its spin derivative, and $I$ is its moment of inertia. Most of the pulsars observed in \gray s \citep[e.g.,][]{Abdo:2009Sci,Fermi-LAT:2013svs,Pletsch:2013iva,Fermi-LAT:2023zzt} have high spin down luminosities ($\dot{E} \gtrsim 10^{33}$\,erg s$^{-1}$). 
This includes both young pulsars (mostly found at low Galactic latitudes) and millisecond pulsars (MSPs).
Comparison of the observed \gray\ luminosity $L_\gamma$ to the spindown luminosity $\dot{E}$ indicates that \gray\ emission accounts for a large fraction of the energy radiated away by many pulsars \citep[with this fraction $\eta \gtrsim 10\%$;][]{Fermi-LAT:2013svs}.
On the other hand, comparing the radio luminosity of pulsars to their spindown luminosities suggests that only a small fraction of the energy is radiated away in radio wavelengths. The radio emission efficiency $\xi$, defined as the fraction of rotational energy transformed into radio emission
\begin{equation}
    \xi = \frac{L_{\textrm{r}}}{\dot{E}},
    \label{eq:xi}
\end{equation}
where $L_{\textrm{r}} (\textrm{erg s}^{-1}) \approx 7.4 \times 10^{27} L_{\textrm{r}}$ (mJy), will always be smaller than 1\% \citep{Szary2014}.
Around half of the pulsars detected in \gray s are also observed in radio, whilst the other half (typically referred to as \textit{radio quiet pulsars}) are not \citep[][]{Fermi-LAT:2008wkm,Abdo:2009Sci,Fermi-LAT:2013svs,Fermi-LAT:2023zzt}. 
The distinction between these two populations likely depends on whether their radio emission is beamed towards the Earth \citep[][]{Abdo:2009Sci}.

The current sample of \gray\ detected pulsars is around 300 \citep[][]{Fermi-LAT:2013svs,Smith:2019ApJ,Fermi-LAT:2023zzt}\footnote{A list of \textit{Fermi} detected pulsars is available at \url{https://confluence.slac.stanford.edu/display/GLAMCOG/Public+List+of+LAT-Detected+Gamma-Ray+Pulsars}} and is limited by the sensitivity of \textit{Fermi}-LAT.
There is an apparent `death-line' around a spindown luminosity of $\dot{E} \sim 10^{33}$\,erg\,s$^{-1}$, which may be a result of the limited sensitivity \citep[][]{Fermi-LAT:2013svs}. 
An extensive search of \gray\ pulsations from more than one thousand pulsars \citep{Smith:2019ApJ} provides a similar result that the lack of detection of pulsars with lower \Edot\ might be due to the limitation of the current instruments, and suggests population synthesis work to explore this specific subpopulation of pulsars. 

Recently, \citet{Song2023} performed a stacking analysis of 12 years of \textit{Fermi} data, targeting the positions of 362 known radio pulsars with no previous \gray\ detections.
In order to reduce the impact of the Galactic \gray\ background, they selected pulsars that lie at high galactic latitudes ($|b| > 20^\circ$).
This probed a population of relatively old, low $\dot{E}$ pulsars, different to the usual population of \gray\ loud pulsars.
\citet{Song2023} report the significant detection of \gray s from this set of pulsars compared to the expected background.
\citet{Song2023} argue that this suggests that the usual \gray\ death-line is predominantly a result of the sensitivity of \textit{Fermi}.
In this paper we attempt to explore the implications of this detection for massive binary and pulsar evolution.

Theoretical population synthesis studies can be used to predict the population of observable pulsars under a set of assumptions.
There have been several previous population synthesis studies of the \gray\ pulsar population.
For example, \citet{Gonthier:2002ApJ} used Monte Carlo simulations to predict the radio and \gray\ pulsar populations that would be observed by {\it Fermi} in the context of the polar cap model for \gray\ emission.
In a seminal work, \citet{Faucher-Giguere:2006ApJ} performed a population synthesis of the Galactic radio pulsar population. It accounted for almost all of the relevant effects, including radio selection effects, magnetic field decay and pulsar orbits in the Galaxy. 
Assumptions were made on distributions of pulsar birth spin period and birth magnetic field strength.
Since then, there have been a number of similar studies building on this work \citep[e.g.,][]{Takata2011,Dirson2022} that include radio and \gray\ selection effects and pulsar evolution theories. 
Modern pulsar population synthesis studies \citep[e.g.,][]{Gullon:2014MNRAS,ShiNg:2024ApJ,Graber:2024ApJ,PardoAraujo:2025A&A} make use of updated assumptions regarding the birth distributions, magnetic field decay, and pulsar emission geometry, and some of them use sophisticated machine learning inference techniques to compare model predictions to observations.

Typically, existing pulsar population synthesis works, such as those done by using \texttt{PsrPopPy} \citep{Bates:2014MNRAS},  do not model the details of the full evolution of the binary stellar systems. 
Neutron stars---and hence pulsars---originate from massive stars, and the majority of massive stars are born in binaries \citep[e.g.,][]{Sana:2012Sci,MoeDiStefano::2017ApJS}. 
Hence, whilst the majority of ``normal", young pulsars (with $P \gtrsim 0.1$\,s) are isolated, most of these objects are expected to have been born in binaries, which are subsequently disrupted by the supernova that formed the neutron star \citep[e.g.,][]{BrandtPodsiadlowksi:1995MNRAS,Kalogera:1996ApJ,Renzo:2019A&A}.
Hence, the binary fraction and spatial distributions (including scale height) of pulsars encode information about their birth locations and natal kicks. 
It is then important to account for a full picture of stellar and binary evolution for a better understanding of the Galactic pulsar population. 
Binary population synthesis code suites, such as COMPAS \citep{COMPAS}, are ideally suited for this task, as they are able to rapidly simulate the evolution of large populations of massive binary stars, incorporating all relevant processes such as natal kicks and pulsar evolution.
Understanding the properties of the Galactic pulsar population across the radio and \gray\ wavelengths will aid in constraining pulsar emission physics and the formation and evolution of massive binary stars.

Some previous studies have combined pulsar population synthesis with binary stellar evolution.
The study of \citet{Kiel:2008MNRAS} is closest to ours in terms of methodology.
They use binary population synthesis to study the Galactic pulsar population, but they focus on radio pulsars and do not explore the \gray\ pulsar population.
More recently, \citet{Titus:2020MNRAS} used binary population synthesis to study the radio pulsar population in the Small Magellanic Cloud.
In this work, we attempt to explore the canonical radio and \gray\ pulsar populations produced through massive binary stellar evolution and compare them to the pulsar populations observed in the Milky Way.
In a follow-up work, we will attempt to understand the full effects on pulsar populations from stellar evolution physics.

As a first study of the Galactic \gray\ pulsar population using COMPAS, one of the main purposes of this work is to verify the binary and pulsar physics included in COMPAS.
This work and its follow-ups can also be used to make predictions for the population of \gray\ pulsars that will be observed in the future with continued operation of {\it Fermi}-LAT.
Other interesting binary pulsar systems include neutron star X-ray binaries \citep{1998Natur.394..344W}, spider pulsars \citep{1988Natur.333..237F} and double neutron star binaries \citep[e.g.,][]{Hulse:1975ApJL,PhysRevLett.119.161101}, and these events can only be studied through the inclusion of binary physics.
In follow-up studies, we will investigate known peculiar binary systems that contain at least one neutron star using COMPAS and other suitable tools.

The remainder of this paper is structured as follows.
In Section~\ref{sec:methods} we describe our methodology.
We present our results in Section~\ref{sec:results}.
Section~\ref{sec:future} focuses on follow-up studies from this work, and we conclude in Section~\ref{sec:conclusions}.

\section{Methods}
\label{sec:methods}

In this section we introduce the methods and theory used in this study, including evolution theory of isolated pulsars, radio observation and selection criteria, and \gray\ observation and selection criteria. 
We also introduce the basics of COMPAS relevant to synthesising pulsar populations.

\subsection{COMPAS}
\label{subsec:compas} 

We simulate a population of massive binary stars using the rapid binary population synthesis suite COMPAS \citep{Stevenson:2017tfq,gomez2018,COMPAS,Mandel:2025COMPAS}. 
COMPAS is based on the Single Star Evolution \citep[SSE,][]{SSE} code and the Binary Star Evolution \citep[BSE,][]{BSE} code with updates and modifications made as described in \citet{COMPAS}. 
Binary interactions such as mass transfer and common envelope evolution are included, but we neglect the effects of tides \citep[][]{COMPAS}.

We use default binary property parameters for our COMPAS simulations \citep{COMPAS}.
These parameters and their values/ranges are based on past theoretical and observational studies.
The mass of the initially more massive star in the binary is drawn from an initial mass function \citep[IMF;][]{Kroupa:2000iv} within the range of 5 to 150 M$_{\odot}$ as we need stars massive enough to leave behind neutron stars at the end of their evolution.
The mass ratio ($q = m_2 / m_1$, $m_2 < m_1$) of the binary is drawn from a uniform distribution between 0.01 and 1, with a minimum secondary mass of $0.1$\,M$_\odot$ \citep[][]{Stevenson:2019ApJ}.
We assume initially circular orbits \citep[][]{BSE}.
We draw the orbital separation from a uniform distribution in log space with a minimum of 0.01 and maximum of 1000 AU \citep{Sana:2012Sci}.
Since we are interested in pulsars formed recently in the Milky Way, we assume a representative value of Galactic metallicity $Z_\mathrm{Gal} = 0.014$ \citep{Asplund:2009ARA&A}.
A list of default COMPAS option values, including mass transfer rate and common envelope evolution, can be found in the link in footnote\footnote{\url{https://github.com/FloorBroekgaarden/Double-Compact-Object-Mergers/blob/main/otherFiles/Table\_with\_detailed\_binary\_\_population\_synthesis\_simulation\_settings.pdf}}.

We assume a binary fraction of 100\% for massive stars and neglect single stars. 
This is appropriate for massive O stars ($M > 16$\,M$_\odot$). For example, \citet{Sana:2012Sci} find that 70\% of massive stars are in close binaries with $\log(P_\mathrm{orb}/\mathrm{d}) < 3.5$, and the binary fraction reaches 100\% when including wider binaries \citep[][]{Sana:2014ApJS} which we include in our COMPAS models \citep[see also discussion in][]{deMinkBelczynski::2015ApJ}. Many of these stars are effectively single, in the sense that they do not interact with their wide binary companion.
Assuming a binary fraction of 100\% is likely an overestimate for B-type stars, which are the progenitors of most neutron stars and have more typical binary fractions of 70\% \citep{MoeDiStefano::2017ApJS}, see also discussion in \citet{Zapartas:2017A&A}. 
However, this effect is likely subdominant to many of the other approximations in this work. We leave a thorough exploration of the binary fraction as a function of mass to future work. 

In this work, neutron stars formed through accretion-induced supernovae, and neutron stars undergoing any recycling/rejuvenation processes are omitted.
Only neutron stars formed through electron-capture supernovae that are not in wide, non-interacting binaries \citep[][]{Willcox:2021kbg}, or through core collapse supernovae are included in the following analysis. The reasoning for this decision is discussed in \S~\ref{subsec:neutron_star_kicks}.  

\subsection{Neutron star kicks}
\label{subsec:neutron_star_kicks}

Neutron stars receive natal kicks at birth \citep[][]{Lyne:1994Nature,Hobbs:2005MNRAS,Verbunt:2017A&A,Igoshev:2020MNRAS,Igoshev:2021MNRAS,DisbergMandel:2025}.
It is common in the pulsar population synthesis literature to use a distribution inferred from the velocities of isolated pulsars as the kick distribution \citep[e.g.,][]{Faucher-Giguere:2006ApJ,Graber:2024ApJ}. 
A common choice is the distribution from \citet{Hobbs:2005MNRAS}\footnote{though see \citet{DisbergMandel:2025} who argue that the Maxwellian distribution from \citet{Hobbs:2005MNRAS} overestimates the typical neutron star velocity due to an error.}.
However, as we are simulating binary progenitors of pulsars, we need to simulate the underlying kick distribution, not the observed kick distribution. 
Several classes of supernovae are expected to give rise to reduced kicks, including electron-capture supernovae \citep[][]{GessnerJanka:2018ApJ}, ultra-stripped supernovae \citep[e.g.,][]{Tauris:2015MNRAS,Suwa:2015MNRAS,Muller:2019MNRAS} and low-mass iron core-collapse supernovae \citep[][]{Stockinger:2020MNRAS,Muller:2019MNRAS}. 

We use the physical remnant prescription from \citet{Mandel:2020qwb} to relate the remnant mass and kick of a neutron star to the properties of its progenitor.
This prescription is based on semi-analytic supernova models, and is calibrated to the Galactic pulsar population \citep[][]{Kapil:2022blf}; that is, the velocity distribution of young isolated pulsars matches those inferred from very long baseline interferometry observations \citep[][]{Deller:2019ApJ}.
Specifically, we use $v_\mathrm{NS} = 520$\,km/s for the scaling pre-factor for neutron star kicks, and $\sigma_\mathrm{NS} = 0.3$ for the fractional stochastic scatter in the kick velocities.
\citet{Kapil:2022blf} find the distribution of kicks in single and binary neutron stars to be similar, so we opt to use a single distribution for both.
As discussed by \citet{Mandel:2022sxv}, neutron star kicks inferred from young pulsars may be overestimated by $\sim 15$\,\% if pulsar beams are narrow and preferentially aligned with the kick direction. The \citet{Mandel:2020qwb} prescription is applied equally to both core-collapse supernovae and electron-capture supernovae.
We do not allow for electron-capture supernovae to occur in wide, non-interacting binaries, to avoid overproducing low-speed pulsars which are not observed \citep[][]{Willcox:2021kbg}.
This choice is also motivated theoretically as binary interactions are expected to widen the parameter space for the production of electron-capture supernovae compared to single stars \citep[e.g.,][]{Podsiadlowski:2004ApJ}.

\subsection{COMPAS Implementation of Pulsar Evolution}
\label{subsec:pulsar_evolution}

This study focuses on canonical pulsars, hence we only describe the evolution of spin and magnetic field of pulsars that are not experiencing any mass transfer.
Here we follow magnetic braking for an isolated pulsar to spin down, with braking index $n = 3$.
The pulsar evolution theory implemented in COMPAS is described in \citet{Chattopadhyay2020}, and is briefly introduced below.

The surface magnetic field strength of isolated pulsars may decay over time. 
The exact timescale on which this decay occurs (if at all) is uncertain (see further discussion in Section~\ref{subsec:free_params}).
The field decay is thought to be due to the Hall effect leading to a redistribution of the magnetic energy, leading to enhanced Ohmic dissipation \citep{Romani:1990Natur, Konar:1997MNRAS.284..311K, Konar:1999MNRAS.303..588K, Konar:1999MNRAS.308..795K,PonsGeppert:2007PRL,Aguilera:2008A&A}.
In COMPAS, for simplicity we assume an exponentially decaying magnetic field of a neutron star 
similar to many recent studies \citep{Kiel:2008MNRAS,Chattopadhyay2020, Dirson2022, ShiNg:2024ApJ}.
The decay of the magnetic field after it spins down for a time duration of $t$
\begin{equation}
    B_f = (B_i - B_{\textrm{min}}) \exp(-t / \tau) + B_{\textrm{min}},
	\label{eq:isolate_B}
\end{equation}
where $B_i$ is the magnetic field strength of the neutron star before it is evolved by time duration $t$ and $B_f$ the magnetic field strength afterwards, $B_{\textrm{min}}$ is the minimum magnetic field of a pulsar which is set to be $10^{8}$\,G \citep[][]{Zhang:2006MNRAS} for the entire analysis, and $\tau$ is the time scale of pulsar magnetic field decay. 

The rate of change of the spin period of a neutron star over time is known as the spin period derivative, \Pdot\ . 
Assuming spindown due to a rotating magnetic dipole in a vacuum,  \Pdot\ is given by
\begin{equation}
    \dot{P} = \frac{8 \pi ^ 2  R ^ 6 B^2 \sin^2{\alpha}}{3 c^3 I P} ,
    \label{eq:isolate_Pdot}
\end{equation}
where $R$ is the radius of the pulsar (we assume $R = 10$\,km for all pulsars at all times in this work), $c$ is speed of light, $\alpha$ is the alignment angle between the magnetic and rotational axes, and
\begin{equation}
    I = (0.237 \pm 0.008) M R^2 [1 + 4.2\frac{\textrm{M}}{\textrm{M}_{\odot}} \frac{\textrm{km}}{\textrm{R}} + 90(\frac{\textrm{M}}{\textrm{M}_{\odot}}\frac{\textrm{km}}{\textrm{R}})^4] ,
    \label{eq:moment_of_inertia}
\end{equation}
being the moment of inertia of the pulsar \citep{lattimer2005}. 
Substituting $B$ from Eq.~\ref{eq:isolate_B} into Eq.~\ref{eq:isolate_Pdot} and integrating over time $t$ gives
\begin{equation}
    P^2 = \frac{16 \pi ^ 2  R ^ 6 \sin^2{\alpha}}{3 c^3 I}  [B_{\textrm{min}}^2 t - \tau B_{\textrm{min}} (B_f - B_i) - \frac{\tau}{2}(B^2_f - B^2_i)] + P_0^2 .
    \label{eq:isolate_P}
\end{equation}
$P_0$ is the spin period of the pulsar at the beginning of the evolved time duration $t$, and $B_i$ and $B_f$ are the pulsar surface magnetic field at the beginning and end of the time step as described in Eq.~\ref{eq:isolate_B}. 
The evolution of $P$ is dependent on $\alpha$ (Eq.~\ref{eq:isolate_Pdot}). 
As pointed out in other pulsar population synthesis studies \citep[e.g.,][]{Dirson2022}, the evolution of $\alpha$ is also dependent on $B$, as the field decay impacts the evolution of $\alpha$.
A simplification in COMPAS is made so that $\alpha$ is set constant and not evolved. This decouples the evolution between $P$/$B$ and $\alpha$.
Some works have suggested that $\alpha$ tends to decrease as the pulsar ages, and the old pulsar populations have near-alignment between the magnetic field and rotational axes \citep{Young:2010MNRAS}. However, results from \citet{Johnston2020} suggest that large $\alpha$ is needed to model both \gray\ and radio pulsars well, especially for younger populations. Hence in this work we set $\alpha = \pi/2$.
Evolution of alignment will be included in a future version of COMPAS.

\subsection{Free Parameters of COMPAS Runs}
\label{subsec:free_params}

In addition to the default binary parameters, we need to propose distributions of pulsar properties at birth for subsequent modelling. 
Previous pulsar population studies \citep[e.g.,][]{Faucher-Giguere:2006ApJ,Popov:2012Ap&SS,Szary2014, 
Johnston2020, Chattopadhyay2020, Dirson2022, Igoshev:2022MNRAS, Du:2024ApJ} have proposed and refined distributions of pulsar properties at birth, such as the surface magnetic field strength and spin period. 
In this work, we also choose the distributions of these two quantities at birth, as well as magnetic field decay timescale ($\tau_d$) to be free parameters. 

By default, COMPAS sets the birth distribution of NS magnetic field as a uniform distribution between $10^{10}$ G to $10^{13}$ G, and the birth distribution of NS spin period as a uniform distribution between 10 to 100 ms. 
These distributions are in general agreement with those observed from young pulsar populations \citep{Manchester:2004bp}.
We exclude extremely high magnetic field strengths with B $> 10^{15}$\,G in the magnetar regime as magnetars are not powered by rotation like typical pulsars, and may have also have a different formation history. 

In our default model, we assume that the magnetic field decay time scale $\tau_d = 500$\,Myr \citep{Chattopadhyay2021}. 
We choose $\tau_d$ as a free parameter since there is no clear consensus on whether the magnetic field of a pulsar decays at all, or how long it takes to decay. 
For example, semi-analytical studies \citep{1970ApJ...160..979G, 1981JApA....2..315V, 1985MNRAS.213..613L} found that the magnetic field decays on the timescale of the order of a million years, with \citet{Dirson2022} recently finding $\tau_d = 4.6\times10^{5}$ yr. 
In comparison, other works \citep{1977ApJ...215..885T, 1987A&A...178..143S, Kiel:2008MNRAS} have shown the opposite, finding that $\tau_d > 100$\,Myr is needed in order for theoretical models to match the observed catalogue. 
Recent work incorporating more sophisticated models of magnetic field decay, pulsar beaming and magnetic field alignment find $\tau_d$ in the range 1--10\,Myr \citep[e.g.,][]{Graber:2024ApJ,ShiNg:2024ApJ}.
Given the uncertainty, we let $\tau_d$ vary from 1\,Myr to 10,000\,Myr in some of our models. 

We list the details of all the models with different pulsar birth distributions and parameters in Table~\ref{tab:compas_parameter}. 
The initial model (INIT) is as described above. Following that, we vary our model parameters in such ways that we cover different ranges, and in some cases different shapes of the distribution. 
Models D1--5, combined with INIT, are specifically targeted to examine the impact of the $\tau_d$ on the results of the simulations, as only $\tau_d$ is varied in models D1--5 compared to INIT. 
For B1 and B2 models, only the range of the birth magnetic field distributions is changed compared to INIT. 
Similarly for P1, P2 and P3 models, only the ranges of the birth spin period distributions are changed and the others are kept the same. 
For PN1, PN2 and PN3 models, the birth spin period distribution is changed to a normal distribution (hence, PN models as in period-normal models) with various means and standard deviations. Model PLN1 gives the birth spin period distribution in a log uniform distribution.
In the BLN series of models, the birth magnetic field strength distribution is changed to a log normal distribution, with different means and standard deviations in models BLN1--4. 
Models P2-D1--5, PN1-B1-D1--5 and PN5-B1-D1--5 are set to explore the impact of $\tau_d$ on different initial distributions of spin period and magnetic field, whilst also varying $\tau_d$ in the same way as in models D1--5.
The remaining three models, BLN1-PN1, BLN4-PN1, and BLN1-PN5, have a combination of the magnetic field distribution from the BL models and the spin period distributions of the PN models, respectively. 
The original form of the log-normal distribution of the birth magnetic field strength is taken from \citet{Szary2014}. 
The BLN1-PN5 model is intended to reproduce the model described in \citet{Szary2014}. 
The birth magnetic field and spin period distributions in BLN5-PLN2-D4 are found in \citet{Graber:2024ApJ}. The birth distributions for BLN6-PLN3-D4 are found in \citet{PardoAraujo:2025A&A}. 
In \citet{Graber:2024ApJ, PardoAraujo:2025A&A} the authors model the magnetic field decay based on numerical simulations of magnetothermal evolution in the
neutron star crust \citep{Vigano:2021} up to 1 Myr, and with a power law decay model beyond 1 Myr. 
Their results are equivalent to a short decay timescale. 
Hence for these two models, the magnetic field decay time scale is set to 1 Myr.
In each COMPAS model, 300,000 binary systems are modelled.

\begin{table*}
	\centering
	\caption{Description of models used for COMPAS simulations in this work. }
	\label{tab:compas_parameter}
	\begin{tabular}{lccccccr} 
		\hline
		Model & $B_{\textrm{birth}}$ Range (G) & $B_{\textrm{birth}}$ Distribution & $P_{\textrm{birth}}$ Range (ms) & $P_{\textrm{birth}}$ Distribution & $\tau_d$ (Myrs) \\
		\hline 

		Initial & ($10^{10}$ - $10^{13}$) & Uniform & (10 - 100) & Uniform & 500 \\
            D1 & ($10^{10}$ - $10^{13}$) & Uniform & (10 - 100) & Uniform & 100 \\
            D2 & ($10^{10}$ - $10^{13}$) & Uniform & (10 - 100) & Uniform & 10 \\
            D3 & ($10^{10}$ - $10^{13}$) & Uniform & (10 - 100) & Uniform & 1,000 \\
            D4 & ($10^{10}$ - $10^{13}$) & Uniform & (10 - 100) & Uniform & 1 \\
            D5 & ($10^{10}$ - $10^{13}$) & Uniform & (10 - 100) & Uniform & 10,000 \\
            
            B1 & ($10^{11}$ - $10^{13}$) & Uniform & (10 - 100) & Uniform & 500 \\
            B2 & ($10^{10}$ - $10^{12}$) & Uniform & (10 - 100) & Uniform & 500 \\
            
            P1 & ($10^{10}$ - $10^{13}$) & Uniform & (10 - 1000) & Uniform & 500 \\
            P2 & ($10^{10}$ - $10^{13}$) & Uniform & (1 - 100) & Uniform & 500 \\
            P3 & ($10^{10}$ - $10^{13}$) & Uniform & (1 - 1000) & Uniform & 500 \\
    
            PN1 & ($10^{10}$ - $10^{13}$) & Uniform & $\mu=75$,$\sigma=25$ & Normal & 500 \\
            PN2 & ($10^{10}$ - $10^{13}$) & Uniform & $\mu=50$,$\sigma=25$ & Normal & 500 \\
            PN3 & ($10^{10}$ - $10^{13}$) & Uniform & $\mu=50$,$\sigma=50$ & Normal & 500 \\

            PLN1 & ($10^{10}$ - $10^{13}$) & Uniform & $\mu=10^{1.60}$,$\sigma=10^{1.40}$ & LogNormal & 500 \\

            BLN1 & $\mu=10^{12.6},\sigma=10^{0.55}$ & LogNormal & (10 - 100) & Uniform & 500 \\
            BLN2 & $\mu=10^{12.0},\sigma=10^{0.55}$ & LogNormal & (10 - 100) & Uniform & 500 \\
            BLN3 & $\mu=10^{12.6},\sigma=10^{0.25}$ & LogNormal & (10 - 100) & Uniform & 500 \\
            BLN4 & $\mu=10^{12.0},\sigma=10^{0.25}$ & LogNormal & (10 - 100) & Uniform & 500 \\

            P2-D1 & ($10^{10}$ - $10^{13}$) & Uniform & (1 - 100) & Uniform & 100 \\
            P2-D2 & ($10^{10}$ - $10^{13}$) & Uniform & (1 - 100) & Uniform & 10 \\
            P2-D3 & ($10^{10}$ - $10^{13}$) & Uniform & (1 - 100) & Uniform & 1,000 \\
            P2-D4 & ($10^{10}$ - $10^{13}$) & Uniform & (1 - 100) & Uniform & 1 \\
            P2-D5 & ($10^{10}$ - $10^{13}$) & Uniform & (1 - 100) & Uniform & 10,000 \\

            PN1-B1 & ($10^{11}$ - $10^{13}$) & Uniform & $\mu=75$,$\sigma=25$ & Normal & 500 \\
            PN1-B1-D1 & ($10^{11}$ - $10^{13}$) & Uniform & $\mu=75$,$\sigma=25$ & Normal & 100 \\
            PN1-B1-D2 & ($10^{11}$ - $10^{13}$) & Uniform & $\mu=75$,$\sigma=25$ & Normal & 10 \\
            PN1-B1-D3 & ($10^{11}$ - $10^{13}$) & Uniform & $\mu=75$,$\sigma=25$ & Normal & 1,000\\
            PN1-B1-D4 & ($10^{11}$ - $10^{13}$) & Uniform & $\mu=75$,$\sigma=25$ & Normal & 1 \\
            PN1-B1-D5 & ($10^{11}$ - $10^{13}$) & Uniform & $\mu=75$,$\sigma=25$ & Normal & 10,000 \\

            PN5-B1-D1 & ($10^{11}$ - $10^{13}$) & Uniform & $\mu=300$,$\sigma=150$ & Normal & 100 \\
            PN5-B1-D2 & ($10^{11}$ - $10^{13}$) & Uniform & $\mu=300$,$\sigma=150$ & Normal & 10 \\
            PN5-B1-D3 & ($10^{11}$ - $10^{13}$) & Uniform & $\mu=300$,$\sigma=150$ & Normal & 1,000\\
            PN5-B1-D4 & ($10^{11}$ - $10^{13}$) & Uniform & $\mu=300$,$\sigma=150$ & Normal & 1 \\
            PN5-B1-D5 & ($10^{11}$ - $10^{13}$) & Uniform & $\mu=300$,$\sigma=150$ & Normal & 10,000 \\
            BLN1-PN1 & $\mu=10^{12.6},\sigma=10^{0.55}$ & LogNormal & $\mu=75$,$\sigma=25$ & Normal & 500\\
            BLN4-PN1 & $\mu=10^{12.0},\sigma=10^{0.25}$ & LogNormal & $\mu=75$,$\sigma=25$ & Normal & 500\\
            BLN1-PN5 & $\mu=10^{12.6},\sigma=10^{0.55}$ & LogNormal & $\mu=300$,$\sigma=150$ & Normal & 500\\

            BLN5-PLN2-D4 & $\mu=10^{13.1},\sigma=10^{0.45}$ & LogNormal & $\mu=10^{2}$,$\sigma=10^{3.38}$ & LogNormal & 1\\
            BLN6-PLN3-D4 & $\mu=10^{13.09},\sigma=10^{0.5}$ & LogNormal & $\mu=10^{2.33}$,$\sigma=10^{3.55}$ & LogNormal & 1\\
		\hline
            
	\end{tabular}
        
\end{table*}

\subsection{Binary Birth, Pulsar Death, \& Parameters}
\label{subsec:pulsar_birth_death}

COMPAS begins evolving each binary system at $T = 0$\,Myr.
The first stage of post-processing is to assign a birth time of the binary system that corresponds to the Milky Way star formation history. 
We consider a uniform star formation history of the Milky Way \citep{gomez2018}, enabling us to draw the birth time of the binary systems from a uniform distribution. 
Without recycling, pulsars rapidly spin down, losing their rotational energy to the point that they can no longer sustain detectable radio or \gray\ emission, at which point we define this as the death of pulsar. 
The typical characteristic age of a canonical \gray\ pulsar is relatively small, roughly $10^{5}$\,yrs \citep{Fermi-LAT:2023zzt}. 
The median characteristic age of radio pulsars selected from the catalogue in this study (as described later in \S~\ref{subsec:radio_results}) is $3.1\times10^{6}$\,yrs. 
While this age is higher than typical \gray\ pulsars, it still suggests that a typical radio pulsar should be formed in a recent star formation episode.
For this reason, we assume that all the neutron stars from the simulation that would potentially be \gray\ emitting are formed in binary systems that are born in the recent history of the Milky Way. Readers should note that in general, the characteristic age is a poor primer for the true age of a pulsar when the neutron star is young or when field decay is included and the star is older than 1 Myr. 

The birth time of any given binary system is then drawn from a uniform distribution with a minimum of 12.0\,Gyr and a maximum of 13.0\,Gyr.
We choose to neglect the population of pulsars produced through accretion-induced collapse of white dwarfs as they represent only a small fraction of the population ($\lesssim 10\%$, as indicated by our COMPAS models), are formed on much longer timescales and may potentially be born with different properties to those born in core-collapse supernovae \citep[][]{Bailyn:1990ApJ,Tauris:2013A&A}.

At this point, the time stamp of the binary system in COMPAS output corresponds to an actual point of time in the history of the Milky Way. 
We make the observation of any neutron star at the present day, assuming the age of the Milky Way is 13.0\,Gyr. 
If the time stamp is not exactly at 13 Gyr, then a linear interpolation of the two nearest data points is used as the recorded parameter value. 
COMPAS provides $P$, \Pdot\ , $B$ and mass of the neutron star at the observation, and one can estimate the spin-down luminosity via Equation~\ref{eq:Edot}.

The emission of a pulsar is turned off when it can no longer produce electron-positron pairs in its magnetosphere. 
More realistic modelling of pulsar polar cap have suggested a broad ``valley'' like region in which pulsars cease their emission \citep{Chen:1993ApJ, Beskin:2022MNRAS}, which is starting to be implemented in some recent pulsar population works \citep[e.g.][]{Sautron:2024A&A}. 
In this work, we choose a more simplistic description with a death-line model.
Before recording the parameters of the neutron star, it is essential to check if the pulsar crosses the radio death-line \citep{rudakritter1994} on the $P$-$\dot{P}$ diagram, which is defined by  
\begin{equation}
    \log_{10} \dot{P} = 3.29 \times \log_{10} P - 16.55, 
    \label{eq:dl1}
\end{equation}
and 
\begin{equation}
    \log_{10} \dot{P} = 0.92 \times \log_{10} P - 18.65, 
    \label{eq:dl2}
\end{equation}
where Eq.~\ref{eq:dl1} is commonly referred to as the death-line for canonical pulsars, and Eq.~\ref{eq:dl2} for MSPs. In the simulation, we decide if a pulsar crosses the death-line (and hence becomes unobservable) by comparing its \Pdot\ to the \Pdot\ values of Eqs.~\ref{eq:dl1} \& ~\ref{eq:dl2} given the current spin period of the neutron star. If \Pdot\ of a neutron star from COMPAS is larger than both \Pdot\ values calculated by Eqs.~\ref{eq:dl1} and ~\ref{eq:dl2}, 
then the pulsar has not yet crossed the death-line and may be observable, and its parameters are recorded. Otherwise, it is discarded from the simulation.
There are other pulsar population models that do not need to explicitly include a death-line, e.g., when assuming both radio and \gray\ luminosity are dependent on \Edot\ \citep{ShiNg:2024ApJ,Graber:2024ApJ,PardoAraujo:2025A&A}.
Our treatment of radio selection effects is given in Section~\ref{subsec:radio_selection_effects}, and \gray\ selection effects in Section~\ref{subsec:gamma_ray_selection}.

\subsection{Dynamical Evolution}
\label{subsec:dynamic}

Massive stars, which in later evolutionary stages form neutron stars, are formed within an exponential disk around the Galactic plane. We assume that these massive stars move along their Galactic orbits without any interruption until the change in the evolutionary status of one of the stars in the binary system. 
As described in \S~\ref{subsec:compas}, we assume all the neutron stars are formed in supernovae. 
Upon formation, the stellar remnant receives a natal kick as described in \S~\ref{subsec:neutron_star_kicks} which may or may not disrupt the binary system, in some case leading to two individual stars. 

Upon the supernova event that forms the neutron star, the binary system is then assigned a random location in the Galaxy following a distribution of a thin, exponential disk \citep{Paczynski1990}: 
\begin{equation}
    p_z(|z|) dz = \exp{(-|z|/z_{\textrm{exp}})}\frac{dz}{z_{\textrm{exp}}},
\end{equation}
where $|z|$ is the absolute value of the vertical distance from the Galactic plane, and
\begin{equation}
    p_R(R) dR = a_R \exp{(-R/R_{\textrm{exp}})} \frac{R}{R^2_{\textrm{exp}}} dR,
\end{equation}
where $a_R = 1.0683$ and $R$ is the distance from the Galactic centre on the Galactic plane. 
We set $z_{\textrm{exp}} = 75$\,pc and $R_{\textrm{exp}} = 4.5$\,kpc following \citet{Paczynski1990} and \citet{Dirson2022}. 
\citet{Quintana:2025MNRAS} recently performed a census of massive OB stars (neutron star progenitors) within 1\,kpc of the Sun and found a scale height close to 75\,pc. 
It is worth noting that a recent study of OB stars in Gaia survey \citep{Li:2019ApJ} has suggested a varying scale height with respect to the Galactic radius.
Previous pulsar population synthesis studies have used both larger and smaller scale heights \citep[][]{Gullon:2014MNRAS,ShiNg:2024ApJ}.

Knowing the location of the newly born neutron star, the time at which it is born, and both the magnitude and direction of the natal kick velocity, we can follow its subsequent motion in the Galactic potential.
We use NIGO \citep{nigo2015}, a numerical galactic orbit integration tool, to evolve the motion of binary systems after the supernova within the Galactic potential. 
NIGO describes the Milky Way as a disc galaxy, with the Galactic potential comprised of three components: a bulge, a disk and a dark matter halo. 
The Galactic bulge is described as a Plummer sphere \citep{flynn1996,plummer1911}:

\begin{equation}
    \Phi_b = - \frac{GM_b}{\sqrt{R^2 + z^2 + b_b^2}},
    \label{eq:bulge}
\end{equation}
where $G$ is the gravitation constant, $M_b = 1.0\times10^{10}$M$_{\odot}$ is the mass of the bulge and $b_b = 0.4$kpc is the scale length and $R^2 = x^2 + y^2$. 

The Galactic disc is modelled by an exponential disc formed by the superposition of three Miyamoto-Nagai potentials \citep{MN1975}:
\begin{equation}
    \Phi_d = -\sum_{n=1}^{3} \frac{G M_{d_n}}{\sqrt{R^2 + [a_{d_n}+\sqrt{b_d^2 + z^2}]^2}},
    \label{eq:disk}
\end{equation}
where $M_{d_n}$ are the masses of each disk, $a_{d_n}$ are related to the disk scale lengths of the three disk components (which are listed in Table~\ref{tab:nigo_pars}) and $b_d = 0.2376$ is related to the disk scale height. 

\begin{table}
    \centering
    \caption{Parameters used for the three components of exponential disk}
    \label{tab:nigo_pars}
    \begin{tabular}{c|ccc}
    \hline
              & 1 & 2 & 3\\
      \hline
      $M_d$ (M$_{\odot}$)  &  $9.0335\times10^{10}$ & $-5.9112\times10^{10}$ & $1.0590\times10^{10}$\\
      $a_d$ (kpc)  &  4.5081 & 9.6483 & 1.5136 \\
      \hline
    \end{tabular}
\end{table}

An NFW dark matter halo \citep{NFW1997} is used in the modelling of the Galactic potential:
\begin{equation}
    \Phi_h = -\frac{GM_h}{r}\ln{(1+\frac{r}{a_h})},
    \label{eq:halo}
\end{equation}
where $M_h = 3.3\times10^{12}$M$_{\odot}$ is the mass of the halo component, $a_h = 45.02$ kpc is the length scale and $r^2 = R^2 + z^2$.

\subsection{Radio Selection Effects}
\label{subsec:radio_selection_effects}

The observed radio pulsar population represents only a small fraction of the total pulsar population in the Milky Way, and may be strongly biased due to observational selection effects. In this section we describe our modelling of the relevant radio selection effects.

\subsubsection{Radio Luminosity}
\label{subsubsec:radio_luminosity}

One of the primary factors determining whether a pulsar is observable is its radio luminosity.
Theoretical work on modelling pulsar magnetosphere has suggested a dependence of radio luminosity on spin-down luminosity \Edot\ \citep{lm2004}.
Recent measurements made by the MeerKAT Thousand Pulsar Array have revealed evidence for such correlation, although it is a weak correlation \citep{Posselt:2023MNRAS}.
Some population studies have also found no significant dependence of radio luminosity on $P$, \Pdot\ or \Edot\ of the pulsar itself \citep{Szary2014}.
For simplicity, we opt to assume no correlation between radio luminosity and spin-down luminosity following \citet{Szary2014}.
We assume that the distribution of pulsar radio luminosities of a pulsar (in units of mJy kpc$^2$ in 1400\,MHz band) follows a log-normal distribution \citep{Szary2014}
\begin{equation}
    \log L_{\textrm{1400}} \sim N(0.5, 1.0), -3.0 \leq \log L_{\textrm{1400}} \leq 4.0 
\end{equation}

If a pulsar crosses the ``death-line'', described in \S~\ref{subsec:pulsar_birth_death}, before the observation is made, then the pulsar is not going to be observed. 
As the pulsar ages, its radio emission efficiency $\xi$ as defined in Eq.~\ref{eq:xi} will increase. As per \citet{Szary2014}, for a pulsar to be detectable in radio, $\xi < 0.01$ must be satisfied.

\subsubsection{PSREvolve: Radio Sensitivity}
\label{subsubsec:psr_evolve}

To determine if a pulsar can be observed by a given radio survey, we use the PSREvolve code \citep{Oslowski2021,Chattopadhyay2020} to calculate the radiometer equation \citep{dewey1985, lm2004} and determine the minimum detectable flux, $S_{\textrm{min}}$, at the sky location at a given signal-to-noise ratio $(S/N_{\textrm{min}})$
\begin{equation}
    S_{\textrm{min}} = \beta \frac{(S/N_{\textrm{min}})(T_{\textrm{rec}}+T_{\textrm{sky}})}{G\sqrt{n_p t_{\textrm{int}}\Delta f}} \sqrt{\frac{W_e}{P-W_e}},
    \label{eq:radiometer}
\end{equation}
where $T_{\textrm{rec}}$ and $T_{\textrm{sky}}$ are the temperature of the receiver and the direction in the sky, respectively; $\Delta f$ is the bandwidth of the receiver, $G$ is the gain of the telescope, $n_p$ is the number of polarisations in the detector, $t_{\rm int}$ is the integration time, $W_e$ is the pulse width and $P$ is the period of the pulsar, and $\beta$ is a noise correction factor that increases the noise with digitisation errors and bandpass distortion.

Pulsar detection has been carried out by almost every major radio facility in the world, over a period of more than 50 years, using a variety of different backends. 
This makes it complicated to individually model each pulsar survey that has been conducted. 
However, one can focus on modelling some of the largest, most successful surveys \citep[e.g.,][]{Faucher-Giguere:2006ApJ,Graber:2024ApJ}.
In this work, we choose pulsar survey parameters for Equation~\ref{eq:radiometer} to represent the Parkes Multibeam Pulsar Survey \citep{manchester2001} and previous binary population synthesis work \citep{Chattopadhyay2021} by setting $\beta = 1$,  $S/N_{\textrm{min}} \geq 10$, $T_{\textrm{rec}} = 24 K$, $\Delta f =$ 288 MHz, $G = 0.65$ K/Jy, $n_p = 2$, and $t_{\rm int} = 2100 s$. 
The sky temperature $T_{\rm sky}$ is determined by the location of the pulsar calculated by NIGO as described in \S~\ref{subsec:dynamic}, and is calculated by scaling the values of sky temperature at 408 MHz from the all-sky survey by \citet{Haslam:1982A&AS} to the frequency of Parkes Multibeam survey of 1374 MHz, assuming the sky temperature spectral index of -2.5 \citep{Reich:1988A&A,Hobbs:2004MNRAS,Guzman:2011A&A}. The pulsar period $P$ is determined by the COMPAS simulation.
The Parkes Multibeam Survey discovered around half of the pulsars detected \citep{manchester2001, Lorimer:2006MNRAS, Sengar:2023MNRAS}. 

The expression of $W_e$ below \citep{2003Natur.426..531B} describes how the interstellar medium (ISM) impacts the observed pulses. 
Free electrons in the Galaxy broaden the intrinsic pulses $W_i$ \citep{2002astro.ph..7156C}, and the ISM scatters the pulsar beam. 
\begin{equation}
    W_e^2 = W_i^2 + \tau_{\textrm{samp}}^2 + (\tau_{\textrm{samp}}\frac{\textrm{DM}}{\textrm{DM}_0})^2 + \tau_{\textrm{scatt}}^2,
\end{equation}
where $W_i$ is the intrinsic width of the pulse, $\tau_{\textrm{samp}} = 250 \mu s$ the sampling time of Parkes Multibeam Survey, and $\tau_{\textrm{scatt}}$ ISM scattering times based on a fit with respect to DM from \citet{Bhat:2004ApJ}, DM is the dispersion measure in the direction of the pulsar and DM$_0$ is the diagonal dispersion measure of the survey. 
While the duty cycle of pulsars varies significantly \citep{1985MNRAS.213..613L}, here we assume $W_i/{P} = 0.05$ for all pulsars for simplicity. 
If the radio luminosity of a pulsar is larger than $S_{\textrm{min}} d^2$, then it can be detected by the chosen radio survey. 

Additionally, only pulsars that are located in the regions observable by the Parkes Multibeam survey are selected. Their Galactic latitude (b) and longitude (l) would satisfy

\begin{equation*}
    (|b| \leq 5), \& (0 \leq l \leq 50 \ {\rm or}\ 260 \leq l < 360)
\end{equation*}

\subsubsection{Radio beaming}
\label{subsubsec:radio_beaming}

Pulsars are considered to be observable during the pulsations, which are strongly beamed. 
A pulsar should only be observable when the beamed pulsation crosses the line of sight from the Earth. 
To this end, we should consider the beaming fraction of pulsars in radio, which depends on different emission mechanisms and geometry of the pulsar. 

The radio beaming fraction of a pulsar, $f_r$ \citep{Tauris1998}, is considered to be
\begin{equation*}
    f_r = 0.09 \log{(P/10)}^2 + 0.03,
\end{equation*}
where $P$ is the rotational period of the pulsar in seconds. 
We use a probabilistic rejection scheme to use the beaming direction as an additional selection criterion to determine the detectability of pulsars in radio and \gray\, assuming a uniformly distributed viewing angle.

\subsection{Gamma-ray Selection Effects}
\label{subsec:gamma_ray_selection}
\subsubsection{Gamma-ray Luminosity \& Flux}
\label{subsubsec:gamma_ray_luminosity}

An empirical description of the relationship between the \gray\ luminosity of a pulsar and its physical properties as a fundamental plane has been proposed \citep{Pulsar_FP2019, Pulsar_FP2022}. 
In this work we estimate pulsar \gray\ luminosity using a fundamental plane in four-dimensional (4D) space that is fitted to 
pulsars in the Fourth {\it Fermi}-LAT Source Catalogue \citep[4FGL;][]{4FGL} using 12-Year survey data \citep[4FGL-DR3;][]{4FGL-DR3}, which contains 190 \gray\ detected pulsars with well defined distances in the ATNF catalogue. \citet[][]{Pulsar_FP2022} find that the \gray\ luminosity is given by,
\begin{equation}
    L_{\gamma} = 10^{14.3\pm1.3} E_{\textrm{cut}}^{1.39\pm0.17} B^{0.12\pm0.03}\dot{E}^{0.39\pm0.05},
    \label{eq:L_gamma}
\end{equation}
where $E_{\textrm{cut}}$ is the cutoff energy of a pulsar's \gray\ SED when modelled by a power-law with the exponential cutoff model, $B$ is the surface magnetic field strength, and $\dot{E}$ is the spin-down luminosity. 
Pulsar \gray\ spectral parameters, cutoff energy $E_{\textrm{cut}}$ and power-law index before cutoff $\Gamma$, are determined following Figures 10 \& 11 in \citet{Pulsar_FP2022}. These two figures show correlations between $E_{\textrm{cut}}$ and \Edot\, and $\Gamma$ and \Edot\ respectively when the pulsar has $10^{33}$ erg/s $<$ \Edot\ $< 10^{38}$ erg/s. Within this \Edot\ range, we assume a power-law correlation of $E_{\rm cut}$ (in GeV) with \Edot\, as well as $\Gamma$ with \Edot\, simply by following the trends in Figures 10 \& 11 in \citet{Pulsar_FP2022} with these following correlations
\begin{equation*}
    \log_{10}E_{\rm cut} = \log_{10}0.5 + 0.26 (\log_{10}\dot{E}-33) + U(0.05), 
\end{equation*}
and
\begin{equation*}
    \Gamma = 0.5 + 0.3 (\log_{10} \dot{E} - 33) + U(0.05),
\end{equation*}
where $U(0.05)$ is a random number drawn from a uniform distribution between -0.05 and 0.05. These relations were estimated visually to fit the approximate trends. Outside this \Edot\ range, we assume constants for $E_{\rm cut}$ and $\Gamma$. When \Edot\ $< 10^{33}$ erg/s, we set 
\begin{equation*}
    \log_{10}E_{\rm cut} = \log_{10}0.5 + U(0.05),
\end{equation*}
and 
\begin{equation*}
    \Gamma = 0.5 + U(0.05).
\end{equation*}
When \Edot\ $> 10^{38}$ erg/s, $\log_{10}E_{\rm cut} = 1.0 + U(0.05)$, $\Gamma = 0.5 +U(0.05)$.

We can then determine the pulsar's energy flux by 
\begin{equation*}
    F_{\gamma} = \frac{L_{\gamma} }{4\pi d^2 c_g},
\end{equation*}
where $d$ is the distance of the pulsar to the Earth and $c_g$ is the flux correction factor \citep{Watters2009}, which depends on the emission mechanism of the pulsar as well as geometry and viewing angle. 
Here we combine results from \citet{Johnston2020} and \citet{Petri2011}, and assume that $c_g$ follows a sigmoid function versus \Edot\ with a maximum of 0.8 \citep{Johnston2020} and a minimum of 0.22 \citep{Petri2011}, while turning over at $10^{33}$\LumUnit\ with an added randomisation drawn from a uniform distribution between -0.1 and 0.1.

\subsubsection{Gamma-ray Sensitivity}
\label{subsubsec:gamma_ray_sensitivity}

The 14-year \textit{Fermi}-LAT Source Catalogue provides an all-sky map of the detection threshold for the catalogue\footnote{\url{https://fermi.gsfc.nasa.gov/ssc/data/access/lat/14yr_catalog/detthresh_P8R3_14years_PL22.fits}}. 
By comparing the energy flux of a pulsar in the simulation with the detection threshold at the corresponding sky location, we would be able to determine if the simulated pulsar would be detectable or not.

In reality, pulsars that are already detected in radio surveys first are easier to find in \gray s when compared to \gray\ blind searches \citep[][]{Fermi-LAT:2013svs, Fermi-LAT:2023zzt}. Figure 17 of \citet{Fermi-LAT:2013svs}
clearly indicates that for the detected pulsars in the 2nd {\it Fermi}-LAT pulsar catalogue, pulsars that are already detected in radio in general are easier to detect. In this work, we define $\alpha$ as the factor to raise the sensitivity floor from the default 4FGL-DR4 sensitivity when the pulsar is not detected in the radio; and $\beta$ as the factor to lower the sensitivity floor when the pulsar is detected in the radio. Evaluating Figure 17 in \citet{Fermi-LAT:2013svs}, we choose $\alpha = 3.0$ and $\beta = 1.5$. 

\subsubsection{Gamma-ray beaming}
\label{subsubsec:gamma_ray_beaming}

Gamma-ray pulsars have a beaming effect similar to that of radio pulsars. In \citet{Johnston2020}, a description of young and energetic (with \Edot\ $> 10^{35}$ erg/s) pulsars' \gray\ beaming fraction ($f_g$) is in their Table 3. We adopt this description in determining the beaming fractions of our modelled pulsars. As \citet{Johnston2020} does not provide a beaming fraction for pulsars with \Edot\ $< 10^{35}$ erg/s, here we assume that as \Edot\ decreases, the beaming fraction also decreases. The choices of beaming fraction can be described as

\[
    f_g = 
\begin{cases}
    0.92,& \text{if } \dot{E} > 10^{38}\ {\rm erg\ s}^{-1}\\
    0.82,& \text{if } 10^{37} < \dot{E} \leq 10^{38}\ {\rm erg\ s}^{-1}\\
    0.67,& \text{if } 10^{36} < \dot{E} \leq 10^{37}\ {\rm erg\ s}^{-1}\\
    0.50,& \text{if } 10^{35} < \dot{E} \leq 10^{36}\ {\rm erg\ s}^{-1}\\
    0.40,& \text{if } 10^{33} < \dot{E} \leq 10^{35}\ {\rm erg\ s}^{-1}\\
    0.30,& \text{if } \dot{E} \leq 10^{33}\ {\rm erg\ s}^{-1}
\end{cases}
\]

The assumptions above are made based on observations and outer gap model of young, powerful pulsars \citep{Johnston2020}. According to \citet{2012A&A...545A..42P}, this choice can be dependent on the pulsar emission model and is consistent with what we have chosen.

We also use a probabilistic rejection scheme same as that described in \S~\ref{subsubsec:radio_beaming} similarly for radio pulsars. If beaming is considered in the chosen emission mechanism of the \gray\ pulsar in the simulation, then it is used as an additional selection criterion to determine the detectability of this pulsar in \gray\ , while also assuming a uniformly distributed viewing angle.

\subsection{Reusing COMPAS Output}
\label{subsec:reusing_COMPAS}

Depending on the model parameters, roughly 30\% to 50\% of binary systems would produce at least one neutron star in each COMPAS simulation. However, due to the selection effects introduced above, only an extremely small portion of these simulated neutron stars are recorded as observable. Adhering to the principle of effectively utilising the simulation, we develop a scheme to reuse each model. Each time the simulation is re-used, the birth time and location of a neutron star is re-assigned. We then make the star travel through the Galactic potential described above, and make the observation again at 13 Gyr, with re-assigned radio and \gray\ luminosity. This re-using scheme is applied 9 times to each simulation, and along with the original simulation, it is equivalent to 3,000,000 binary systems simulated in each COMPAS model.

\section{Results \& Discussions}
\label{sec:results}

The results of the COMPAS simulations using the setups from \S~\ref{sec:methods} are presented in this section, as well as the discussion on the implications of these results. The results are compared to the pulsar catalogue to determine the goodness of the different models. This will help us determine the model of pulsar initial conditions that best fits both the radio and \gray\ observations. We will first employ a statistical method to compare the distribution of pulsar physical properties to those reported in the catalogues, then compare the number of different types of pulsars produced in different simulation models. These results are combined to choose a model that best describe the birth distributions of pulsar properties. We further discuss the impact of the results from the selection effects of \gray\ pulsars. We also aim to provide some insights into the \gray\ emission of low \Edot\ pulsars.  

\subsection{Pulsar Birth Rate}
\label{subsec:sfrate}

In this section, we calculate the pulsar birth rate (PBR) in our COMPAS simulations.
Only binaries where the primary (initially more massive) star is within the initial mass range of 5 \Msun\ to 150 \Msun\ are modelled using COMPAS, since lower mass stars are not expected to produce pulsars.
Because of this setup, we need to account for the low-mass end of the initial mass function (IMF) when calculating the pulsar birth rate. 
By default, we use a \citet{Kroupa:2000iv} broken power-law IMF, with a slope of 0.3 between 0.01\,\Msun\ and 0.08\,\Msun\ ; slope of 1.3 between 0.08\,\Msun\ and 0.5\,\Msun\ ; slope of 2.3 between 0.5\,\Msun\ and 200\,\Msun .
To correct for the lower mass stars (0.08 \Msun\ $< M <$ 5.0 \Msun ), we use the {\tt CosmicIntegration} Python package in COMPAS to calculate a correction factor, which is the ratio of the true average mass produced per binary and the averaged mass produced per binary in COMPAS.
Using the implementation of the Kroupa IMF in {\tt CosmicIntegration}, we draw a sample of $10^{9}$ binaries with the mass range between 0.08 \Msun\ to 150 \Msun\ and calculate the average mass per binary to be $\bar{M}_{\rm true} = 1.33$\,\Msun .
Next we calculate the average mass per binary in COMPAS simulations. Since all the COMPAS models in this study use the same parameters for initial mass and orbit for the binaries and the same evolution physics, they all have the same pulsar birth rate.
In each COMPAS simulation, the total mass simulated in binaries within the 5 to 150\,\Msun\ range is $M_{\rm tot, COMPAS} = 8.05 \times 10^6$\,\Msun . 
The average mass formed per binary in one COMPAS simulation is $\bar{M}_{\rm COMPAS} = 20.3$ \,\Msun .
The correction factor $\kappa$ is calculated as $\kappa = \bar{M}_{\rm COMPAS} / \bar{M}_{\rm true} = 15.14$.

With $N_\mathrm{psr} = 1.2 \times 10^5$ neutron stars produced in each simulation, the pulsar yield (the number of pulsars born per Solar mass of star formation) is calculated as,
\begin{equation}
    \zeta_\mathrm{PSR} = N_\mathrm{psr}  / (M_{\rm tot, COMPAS} \times \kappa) = 1.3 \times 10^{-3}\,\mathrm{M}_\odot^{-1} .
    \label{eq:pulsar_yield}
\end{equation}
To translate the pulsar yield into a PBR, we have to assume a star formation history.
The observed star formation rate (SFR) over the most recent history of the Milky Way is around $1.65 \pm 0.19$ \Msun\ yr$^{-1}$ \citep{2015ApJ...806...96L}, $1.9 \pm 0.4$ \Msun yr$^{-1}$ \citep{2011AJ....142..197C} or $3.3^{+0.7}_{-0.6}$ \Msun yr$^{-1}$ \citep{Zari:2023A&A}.
Other works, such as \citet{2022ApJ...929L..18E}, estimate the Milky Way star formation rate in the range of 0.50--5.93\,\Msun\ yr$^{-1}$.
The pulsar birth rate scales linearly with the assumed SFR.
Here we assume the star formation rate in the recent Milky Way history is $\mathrm{SFR} = 1.65$\Msun\,yr$^{-1}$, and the pulsar birth rate is then,
\begin{equation}
    \mathrm{PBR} = 2.1 \times 10^{-3}\, \mathrm{yr}^{-1} \left( \frac{\mathrm{SFR}}{1.65\, \mathrm{M}_\odot \, \mathrm{yr}^{-1}} \right) 
    \left( \frac{\zeta_\mathrm{PSR}}{1.3 \times 10^{-3}\, \mathrm{M}_\odot^{-1}} \right)
    \label{eq:PBR}
\end{equation}
or 1 pulsar born every 466 years.
This value is lower than most of the literature, e.g., 1/50 yr \citep{Keane:2008MNRAS}, 1/46 yr to 1/58 yr \citep{Graber:2024ApJ}, 1/70 yr \citep{Dirson2022}, and 1/100 yr \citep{Johnston2020}. 
The main reason for this difference could be due to some of the formation channels not included in this work, such as accretion induced collapse (AIC), stellar mergers and double white dwarf mergers. Previous COMPAS models \citep[e.g.,][]{Stevenson:2019ApJ} have also found the predicted rate of core collapse supernova (CCSN) to be somewhat low compared to the observed rate.
Increasing the SFR would also increase the pulsar birth rate in our work, with a detailed exploration on this effect described in \S~\ref{subsec:bestfit}.

\subsection{Results: Radio Pulsars}
\label{subsec:radio_results}

We start the discussion of the results of the simulations by comparing the predicted and observed distributions of various physical quantities of the pulsar populations between the catalogue and the simulations. We first discuss radio pulsars, and then discuss \gray\ pulsars in Section~\ref{subsec:gamma_ray_results}. 

For model comparison purposes, we select Galactic radio pulsars that are not associated with any globular clusters, have positive radio flux at 1400 MHz and have an association with Parkes. 
Only pulsars with $P > 0.03$ s and $\dot{P} > 10^{-17}$ s/s are chosen, to ensure that the sample does not include pulsars that are or have been going through binary interactions/recycling. 
All pulsars are chosen from the ATNF Pulsar catalogue v2.5.1.

\begin{figure*}
    \hspace*{-2.5cm}
    \includegraphics[width=\textwidth]{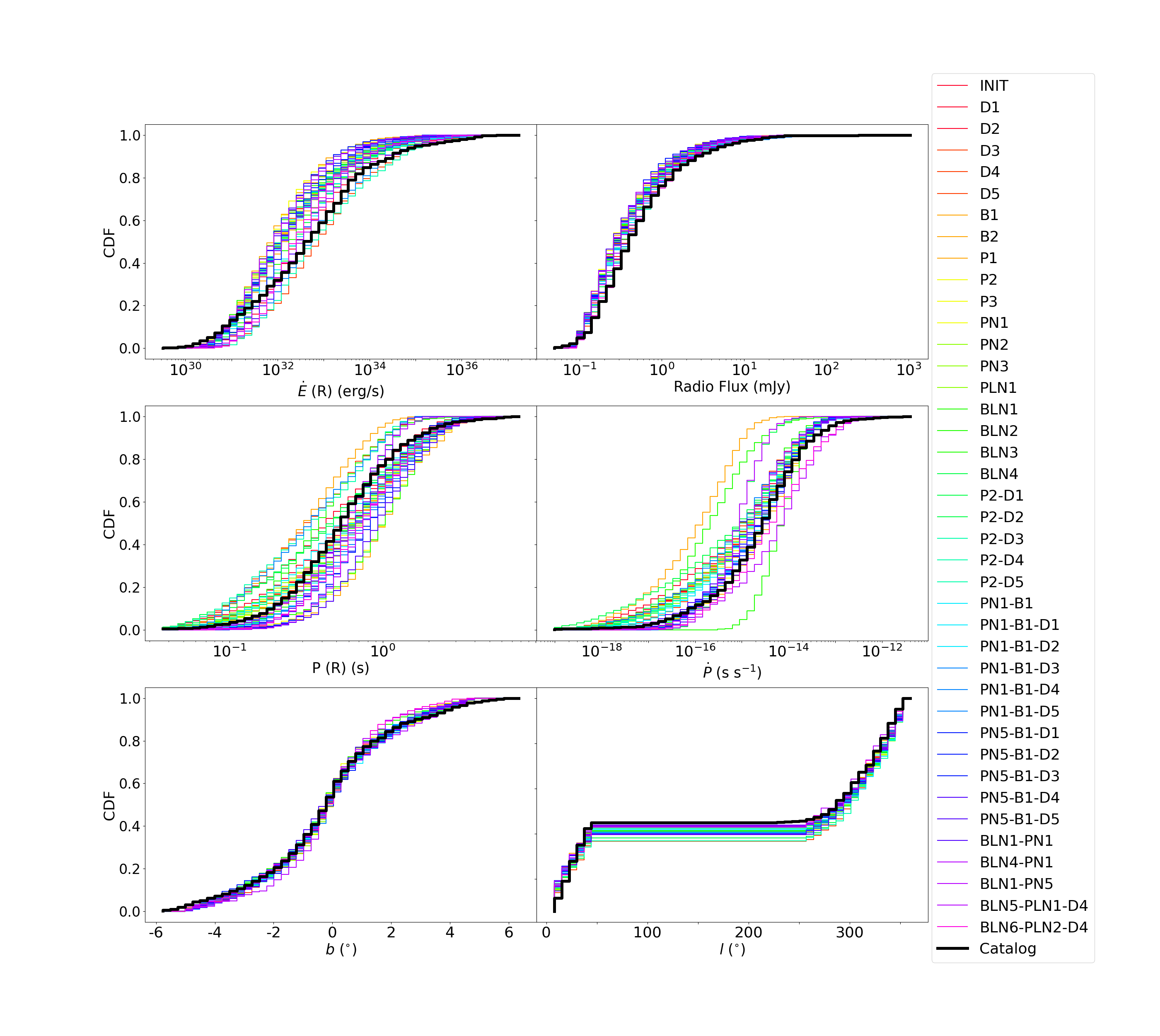}
    \caption{CDFs of physical quantities for radio pulsars produced by the COMPAS models described in Table~\ref{tab:compas_parameter} (coloured), compared to the data from the catalogues (black). CDFs from a given model are drawn from a randomised sampling from 10 realisations. 
    From top left to bottom right, the panels show the distributions of \Edot ,  radio flux, period, \Pdot, Galactic latitude, and Galactic longitude. }
    \label{fig:cdfs_all_models_radio}
\end{figure*}

The cumulative distribution functions (CDFs) for each parameter of radio pulsars for each of the models listed in Table~\ref{tab:compas_parameter} are plotted in Figure~\ref{fig:cdfs_all_models_radio}\footnote{\url{https://github.com/yuzhesong/compas\_pulsar\_paper.git} contains all relevant plots for all models in this study.}. 
Overall, several COMPAS models yield distributions of most physical quantities that align well with the observations.  
The largest variation is seen in the distributions of $P$ and $\dot{P}$.

We use statistical tests to examine how well the predicted parameter distributions match the observed distributions for each physical parameter of the pulsar populations.
These physical parameters include: radio flux, \gray\ flux, spin period $P$, spin-down rate \Pdot\ , spin-down luminosity \Edot\ , Galactic longitude (l) and latitude (b).
Given that the observed data and simulations have different-sized samples, and the physical parameters for each pulsar are independent measurements that is not influenced by measuring a different pulsar, we use Mann-Whitney U test \citep[MWU test;][]{MWUtest} to make the comparison.
The MWU test is chosen because it is a nonparametric test independent of underlying distributions.
The null hypothesis is that both the predicted and observed samples are drawn from the same underlying distribution.
Given the model and simulation have different size, and there is no certain theory on the shape of the distribution of each parameter, it is important to use a statistical test that is independent of the distribution itself. 
We record the $p$-value of the test for a given physical quantity between the comparison of simulation and catalogue.
If $p < 0.05$, it means we reject the null hypothesis that the two compared samples are drawn from the same distribution.
When $p > 0.05$, we cannot reject the null hypothesis, which means we consider the two distributions to be the same.

\begin{table}
	\centering
	\caption{Columns 2 - 7: P-values of Mann-Whitney U Test, comparing the distribution of physical quantities obtained from simulations and catalogue for radio pulsars. Quantities listed in the second to the second last columns are, respectively, radio flux, period, \Pdot , \Edot , Galactic latitude, and Galactic longitude. 
    The last column: the averaged number of the physical quantities with p-values $> 0.05$ out of all 10 realisations of each model. 
    For a given model, the p-value for each parameter is determined using the random realisation of the model that is shown in Figure~\ref{fig:cdfs_all_models_radio}. }
	\label{tab:pvalues-radio}
	\begin{tabular}{lcccccccc} 
		\hline
		Model  & Flux(R) &  Period(R)& $\dot{P}$(R) & \Edot (R) & b(R) & l(R) & NO \\
		\hline
            INIT & 0.0 & 0.13 & 0.0 & 0.0 & 0.38 & 0.09 & 2.4 \\ 
            D1 & 0.0 & 0.25 & 0.0 & 0.0 & 0.5 & 0.52 & 2.4 \\ 
            D2 & 0.0 & 0.0 & 0.0 & 0.38 & 0.35 & 0.14 & 2.5 \\ 
            D3 & 0.0 & 0.01 & 0.0 & 0.0 & 0.26 & 0.42 & 2.1 \\ 
            D4 & 0.24 & 0.0 & 0.0 & 0.0 & 0.14 & 0.01 & 1.7 \\ 
            D5 & 0.0 & 0.0 & 0.0 & 0.0 & 0.35 & 0.14 & 2.0 \\ 
            B1 & 0.0 & 0.0 & 0.06 & 0.0 & 0.71 & 0.58 & 2.3 \\ 
            B2 & 0.0 & 0.0 & 0.0 & 0.0 & 0.29 & 0.18 & 2.0 \\ 
            P1 & 0.0 & 0.0 & 0.15 & 0.0 & 0.4 & 0.5 & 2.9 \\ 
            P2 & 0.0 & 0.04 & 0.0 & 0.0 & 0.14 & 0.06 & 1.9 \\ 
            P3 & 0.0 & 0.0 & 0.57 & 0.0 & 0.74 & 0.48 & 2.9 \\ 
            PN1 & 0.0 & 0.0 & 0.0 & 0.0 & 0.48 & 0.07 & 1.6 \\ 
            PN2 & 0.0 & 0.09 & 0.0 & 0.0 & 0.54 & 0.6 & 2.6 \\ 
            PN3 & 0.0 & 0.0 & 0.0 & 0.0 & 0.09 & 0.34 & 2.0 \\ 
            PLN1 & 0.0 & 0.0 & 0.0 & 0.0 & 0.23 & 0.13 & 2.0 \\ 
            BLN1 & 0.0 & 0.0 & 0.0 & 0.0 & 0.84 & 0.48 & 1.6 \\ 
            BLN2 & 0.0 & 0.0 & 0.0 & 0.0 & 0.79 & 0.19 & 1.9 \\ 
            BLN3 & 0.0 & 0.0 & 0.0 & 0.0 & 0.88 & 0.06 & 1.9 \\ 
            BLN4 & 0.0 & 0.56 & 0.0 & 0.0 & 0.47 & 0.21 & 2.9 \\ 
            P2-D1 & 0.0 & 0.0 & 0.0 & 0.0 & 0.75 & 0.02 & 1.6 \\ 
            P2-D2 & 0.0 & 0.0 & 0.0 & 0.01 & 0.74 & 0.14 & 2.6 \\ 
            P2-D3 & 0.0 & 0.0 & 0.0 & 0.0 & 0.24 & 0.0 & 1.8 \\ 
            P2-D4 & 0.46 & 0.0 & 0.0 & 0.0 & 0.8 & 0.72 & 2.9 \\ 
            P2-D5 & 0.0 & 0.01 & 0.0 & 0.0 & 0.39 & 0.88 & 2.0 \\ 
            PN1-B1 & 0.0 & 0.0 & 0.23 & 0.0 & 0.47 & 0.05 & 2.3 \\ 
            PN1-B1-D1 & 0.0 & 0.0 & 0.0 & 0.0 & 0.57 & 0.36 & 2.0 \\ 
            PN1-B1-D2 & 0.0 & 0.41 & 0.0 & 0.0 & 0.25 & 0.57 & 3.0 \\ 
            PN1-B1-D3 & 0.0 & 0.0 & 0.16 & 0.0 & 0.07 & 0.24 & 2.7 \\ 
            PN1-B1-D4 & 0.2 & 0.0 & 0.0 & 0.02 & 0.54 & 0.62 & 3.0 \\ 
            PN1-B1-D5 & 0.0 & 0.0 & 0.11 & 0.0 & 0.4 & 0.29 & 2.7 \\ 
            PN5-B1-D1 & 0.0 & 0.0 & 0.07 & 0.0 & 0.18 & 0.84 & 1.9 \\ 
            PN5-B1-D2 & 0.0 & 0.0 & 0.01 & 0.0 & 0.25 & 0.56 & 2.8 \\ 
            PN5-B1-D3 & 0.0 & 0.0 & 0.92 & 0.0 & 0.83 & 0.97 & 2.8 \\ 
            PN5-B1-D4 & 0.23 & 0.14 & 0.04 & 0.07 & 0.96 & 0.39 & 3.9 \\ 
            PN5-B1-D5 & 0.0 & 0.0 & 0.27 & 0.0 & 0.35 & 0.35 & 3.0 \\ 
            BLN1-PN1 & 0.0 & 0.0 & 0.0 & 0.0 & 0.86 & 0.91 & 1.9 \\ 
            BLN4-PN1 & 0.0 & 0.34 & 0.0 & 0.0 & 0.44 & 0.97 & 3.0 \\ 
            BLN1-PN5 & 0.0 & 0.0 & 0.0 & 0.0 & 0.23 & 0.92 & 1.9 \\ 
            BLN5-PLN2-D4 & 0.21 & 0.0 & 0.0 & 0.74 & 0.22 & 1.0 & 3.4 \\ 
            BLN6-PLN3-D4 & 0.07 & 0.0 & 0.0 & 0.11 & 0.68 & 0.07 & 3.6 \\ 
            \hline
	\end{tabular}
\end{table}

To account for the stochastic variation due to the limited sample size from the models, the observation part of post-processing is repeated 10 times. 
In Table~\ref{tab:pvalues-radio}, the $p$-values of the MWU test are shown for a selection of the physical quantities of the pulsars compared between each model and the catalogue. 
The $p$-values listed in Table~\ref{tab:pvalues-radio} are based on the random realisation of the model shown in Figure~\ref{fig:cdfs_all_models_radio}. 
To further quantify model performance, we introduce the metric \textbf{NO}, defined as the average number of physical quantities (out of six considered) for which the MWU test $p$-value exceeds 0.05 across the 10 realisations.  
A higher NO value indicates better agreement with the observed distributions, with a maximum possible score of 6.
Notable trends and features from Table~\ref{tab:pvalues-radio} are discussed below.

The spatial distributions of the simulated radio pulsars closely match those in the catalogue. This is evident as the $p$-values for $b$ across all models, and for $l$ in all but the D4 and P2-D3 models, are larger than 0.05, meaning the simulation results align well with the catalogue distributions shown in Figure~\ref{fig:cdfs_all_models_radio}.
This is also an indication that the Galactic models are well-suited for representing radio pulsars.

Only six models successfully reproduce the distributions of radio flux: D4, P2-D4, PN1-B1-D4, PN5-B1-D4, BLN5-PLN2-D4, and BLN6-PLN3-D4. Seven models capture the spin period distribution, nine for the \Pdot\ distribution, and four for the \Edot\ distribution.  
Further examination of the CDF plots for these properties indicates that models B1, BLN1, BLN1-PN1, PN1, D1, D5, P2, PN1-B1, PN1-B1-D2, PN3, PN5-B1-D4, BLN5-PLN2-D4, and BLN6-PLN3-D4 visually exhibit a good fit.

Based solely on the MWU test p-values, model PN5-B1-D4 best matches the observed radio population. 
It achieves an average \textbf{NO} score of 3.9, meaning that on average, 3.9 out of 6 physical quantities have p-values exceeding 0.05 across the 10 realisations.  
The next best-performing models are BLN6-PLN3-D4 with NO score of 3.6, BLN5-PLN2-D4 with NO score of 3.4, then BLN4-PN1 and PN1-B1-D4 each with NO score of 3.0.  
Close behind are models PN5-B1-D5, PN1-B1-D2, P2-D4, P3, P1, and BLN4, each with an average of 2.9 quantities showing p-values above 0.05.
Interestingly, despite good visual comparisons, models PN1-B1-D4, BLN4-PN1, P2-D4, P3, P1 and BLN4 do not exhibit a high number of physical quantities p-values exceeding 0.05.  
All the models mentioned earlier in this paragraph present a mixture of long (PN5 series models and P3) and short (PN1 series models) initial spin periods; large (INIT, $\tau_d = 500$ Myr series models) and small (D2/D4 series models) magnetic decay timescale. All the models mentioned earlier in this paragraph favour larger initial magnetic field strengths.

\subsection{Results: $\gamma$-ray Pulsars}
\label{subsec:gamma_ray_results}

\begin{figure*}
    \hspace*{-2.5cm}
    \includegraphics[width=\textwidth]{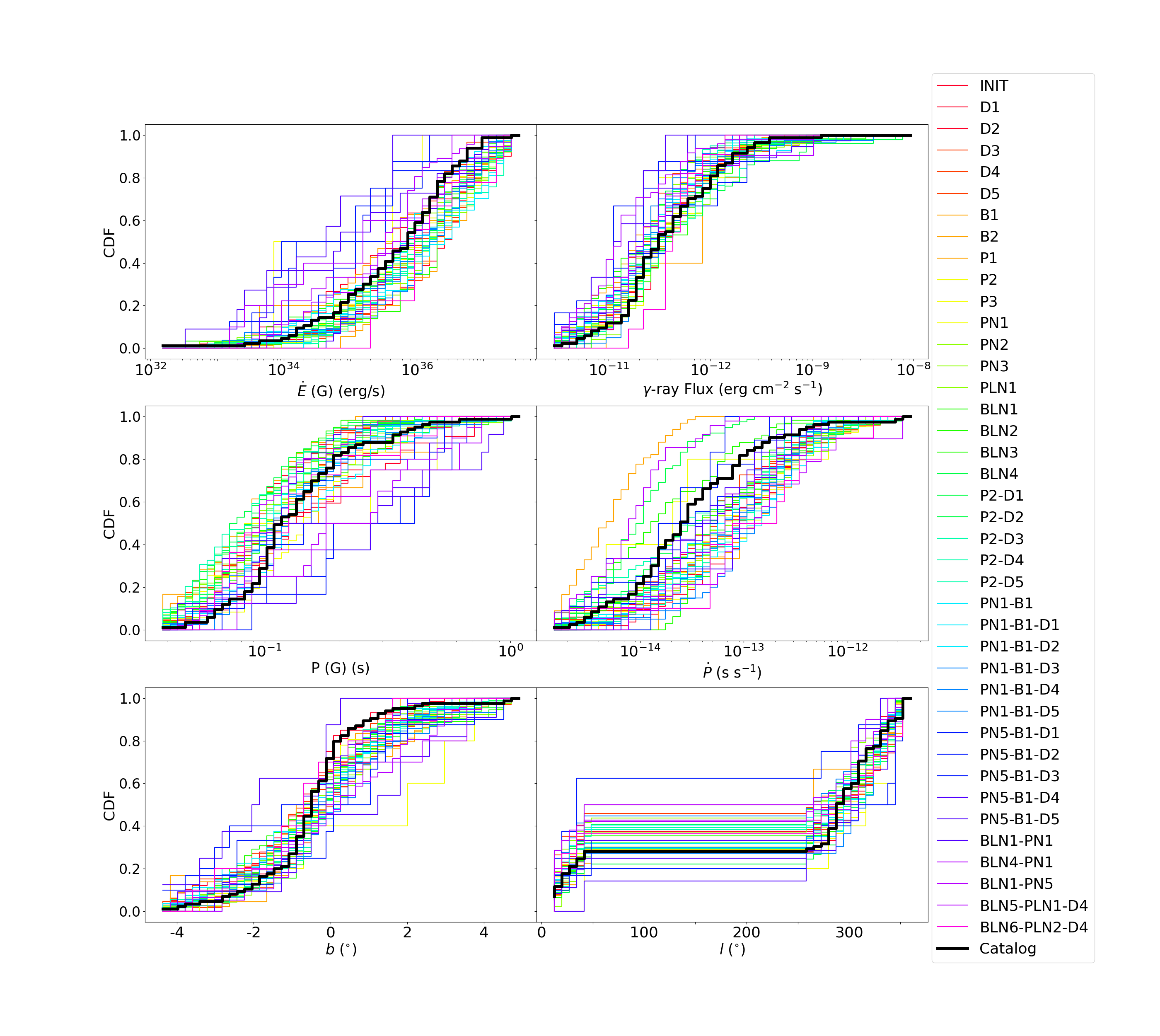}
    \caption{CDFs of physical quantities for \gray\ pulsars produced by the COMPAS models described in Table~\ref{tab:compas_parameter} (coloured), compared to the data from the catalogues (black). CDFs from a given model are drawn from a randomised sampling from 10 realisations. 
    From top left to bottom right, the panels show the distributions of \Edot ,  radio flux, period, \Pdot, Galactic latitude (b), and Galactic longitude (l). }
    \label{fig:cdfs_all_models_gamma}
\end{figure*}

The results of modelling \gray\ pulsars, and the comparisons to the catalogue are presented in this section. 
The catalogue \gray\ pulsars are chosen from 4FGL-DR4, the 14-Year Version of the 4th {\it Fermi}-LAT Source catalogue\footnote{\url{https://fermi.gsfc.nasa.gov/ssc/data/access/lat/14yr_catalog/}} \citep{4FGL-DR4}. Only Galactic \gray\ pulsars not associated with any globular clusters that have $P > 0.03$ s and $\dot{P} > 10^{17}$ s/s are chosen for model comparison.
To ensure consistency, both the catalogue and simulated \gray\ pulsars are selected from the regions covered by the Parkes Multibeam survey. 
The simulated \gray\ pulsars are compared to the catalogue \gray\ pulsars using the same approach applied to radio pulsars, as described in \S~\ref{subsec:radio_results}.  
For the \gray\ pulsars, the same random realisations of each model, as used in \S~\ref{subsec:radio_results}, are selected to plot the CDFs in Figure~\ref{fig:cdfs_all_models_gamma} and to list the MWU test p-values in Table~\ref{tab:pvalues_gamma}.

\begin{table}
	\centering
	\caption{Columns 2 - 7: P-values of Mann-Whitney U Test, comparing the physical quantities obtained from simulations and catalogue for \gray\ pulsars. Quantities listed in the second to the second last columns are, respectively, \gray\ flux, period, \Pdot , \Edot , Galactic latitude, and Galactic longitude. 
    The last column: the averaged number of the physical quantities with p-values $> 0.05$ out of all 10 realisations of each model. 
    For a given model, the p-value for each parameter is determined using the random realisation of the model that is shown in Figure~\ref{fig:cdfs_all_models_gamma}. }
	\label{tab:pvalues_gamma}
	\begin{tabular}{lccccccccc} 
		\hline
		Model  & Flux(G) &  Period(G)& $\dot{P}$(G) & \Edot (G) & b(G) & l(G) & NO \\
		\hline
            INIT & 0.43 & 0.92 & 0.0 & 0.01 & 0.2 & 0.9 & 4.4 \\ 
            D1 & 0.26 & 0.92 & 0.16 & 0.43 & 0.44 & 0.4 & 4.3 \\ 
            D2 & 0.73 & 0.0 & 0.0 & 0.0 & 0.44 & 0.78 & 3.0 \\ 
            D3 & 0.25 & 0.09 & 0.0 & 0.0 & 0.77 & 0.16 & 3.1 \\ 
            D4 & 0.67 & 0.0 & 0.07 & 0.0 & 0.03 & 0.97 & 2.7 \\ 
            D5 & 0.02 & 0.28 & 0.04 & 0.07 & 0.83 & 0.86 & 4.8 \\ 
            B1 & 0.78 & 0.08 & 0.0 & 0.0 & 0.05 & 0.72 & 3.9 \\ 
            B2 & 0.0 & 0.0 & 0.0 & 0.61 & 0.25 & 0.72 & 3.0 \\ 
            P1 & 0.87 & 0.97 & 0.08 & 0.16 & 0.83 & 0.46 & 5.4 \\ 
            P2 & 0.79 & 0.06 & 0.01 & 0.0 & 0.74 & 0.36 & 4.6 \\ 
            P3 & 0.67 & 0.59 & 0.59 & 0.61 & 0.24 & 0.32 & 5.6 \\ 
            PN1 & 0.88 & 0.43 & 0.01 & 0.11 & 0.76 & 0.86 & 4.6 \\ 
            PN2 & 0.58 & 0.03 & 0.0 & 0.0 & 0.36 & 0.33 & 2.9 \\ 
            PN3 & 0.23 & 0.02 & 0.08 & 0.0 & 0.05 & 0.63 & 3.2 \\ 
            PLN1 & 0.64 & 0.0 & 0.17 & 0.0 & 0.88 & 0.06 & 3.2 \\ 
            BLN1 & 0.64 & 0.3 & 0.01 & 0.01 & 0.35 & 0.63 & 3.8 \\ 
            BLN2 & 0.01 & 0.0 & 0.0 & 0.21 & 0.4 & 0.49 & 2.6 \\ 
            BLN3 & 0.54 & 0.08 & 0.0 & 0.0 & 0.1 & 0.85 & 3.3 \\ 
            BLN4 & 0.01 & 0.0 & 0.0 & 0.15 & 0.56 & 0.1 & 2.4 \\ 
            P2-D1 & 0.57 & 0.0 & 0.24 & 0.0 & 0.67 & 0.77 & 3.6 \\ 
            P2-D2 & 0.77 & 0.0 & 0.9 & 0.0 & 0.1 & 0.86 & 3.4 \\ 
            P2-D3 & 0.67 & 0.04 & 0.0 & 0.0 & 0.34 & 0.53 & 3.3 \\ 
            P2-D4 & 0.02 & 0.0 & 0.3 & 0.0 & 0.37 & 0.69 & 3.8 \\ 
            P2-D5 & 0.53 & 0.0 & 0.0 & 0.0 & 0.06 & 0.74 & 2.8 \\ 
            PN1-B1 & 0.28 & 0.4 & 0.0 & 0.0 & 0.96 & 0.43 & 4.0 \\ 
            PN1-B1-D1 & 0.15 & 0.52 & 0.03 & 0.09 & 0.64 & 0.57 & 4.7 \\ 
            PN1-B1-D2 & 0.11 & 0.43 & 0.0 & 0.08 & 0.2 & 0.25 & 3.8 \\ 
            PN1-B1-D3 & 0.92 & 0.57 & 0.0 & 0.0 & 0.03 & 0.31 & 3.0 \\ 
            PN1-B1-D4 & 0.18 & 0.83 & 0.01 & 0.09 & 0.03 & 0.22 & 4.2 \\ 
            PN1-B1-D5 & 0.1 & 0.64 & 0.0 & 0.02 & 0.49 & 0.42 & 3.9 \\ 
            PN5-B1-D1 & 0.03 & 0.03 & 0.76 & 0.09 & 0.67 & 0.62 & 5.0 \\ 
            PN5-B1-D2 & 0.35 & 0.48 & 0.38 & 0.61 & 0.82 & 0.28 & 5.5 \\ 
            PN5-B1-D3 & 0.12 & 0.02 & 0.62 & 0.07 & 0.12 & 0.11 & 5.1 \\ 
            PN5-B1-D4 & 0.0 & 0.02 & 0.4 & 0.05 & 0.08 & 0.85 & 4.8 \\ 
            PN5-B1-D5 & 0.03 & 0.53 & 0.74 & 0.31 & 0.19 & 0.92 & 5.2 \\ 
            BLN1-PN1 & 0.4 & 0.1 & 0.03 & 0.01 & 0.94 & 0.49 & 4.3 \\ 
            BLN4-PN1 & 0.0 & 0.0 & 0.0 & 0.17 & 0.13 & 0.1 & 2.8 \\ 
            BLN1-PN5 & 0.43 & 0.1 & 0.12 & 0.51 & 0.76 & 0.17 & 5.7 \\ 
            BLN5-PLN2-D4 & 0.66 & 0.67 & 0.0 & 0.0 & 0.21 & 0.82 & 4.1 \\ 
            BLN6-PLN3-D4 & 0.22 & 0.95 & 0.0 & 0.01 & 0.47 & 0.27 & 4.2 \\ 
            \hline
	\end{tabular}
\end{table}

The first notable observation is that the simulated \gray\ pulsar populations appear to more closely resemble the catalogue, compared to the modelled radio pulsar populations. 
This is reflected in the greater number of physical quantities for \gray\ pulsars with $p$-values larger than 0.05.
Specifically, 31 models achieve a fit to \gray\ flux with p-values $> 0.05$, while 23 models do so for spin period, 14 for \Pdot\ , and 18 for \Edot\ . 
These values are all higher than those reported for the results from the same models on radio pulsars in Table~\ref{tab:pvalues-radio}.
While it is possible that our \gray\ models are better than the radio ones, another possibility is that the smaller observational sample of \gray\ pulsars and the lower number of \gray\ pulsars produced in these models (see \S~\ref{subsec:bestfit}, especially Table~\ref{tab:number_ns}) make it easier to fit the simulation to observation, resulting in higher $p$-values due to small number statistics.
As with the radio pulsars, most models produce distributions of physical parameters that are close to those of the observation.
However, greater scatter is present in Figure~\ref{fig:cdfs_all_models_gamma}, often attributable to low number statistics. 
Examining Table~\ref{tab:pvalues_gamma} indicates that many of these models provide a good fit to the catalogue.

Based solely on the MWU test p-values, models BLN1-PN5, P3 and PN5-B1-D2 exhibit average of more than 5.5 quantities with $p$-values larger than 0.05. Models P1, PN5-B1-D5, PN5-B1-D3 and PN5-B1-D1 have an average of over 5.0, while models D5, PN5-B1-D4, PN1-B1-D1, P2 and PN1 have averages larger than 4.5, and models INIT, D1, BLN1PN1, PN1B1D4, PN1B1, BLN5-PLN2-D4 and BLN6-PLN3-D4 have averages above 4.0. Once again, similar to the case of radio pulsars, these models cover both long and short initial spin periods, as well as large and small $\tau_d$. 

\subsection{Comparing predicted and observed numbers of pulsars}
\label{subsec:compare_numbers}

Another straightforward way to determine if a COMPAS model is a good fit for the observation is to compare the number of pulsars predicted by the model to those listed in the catalogues.
We evaluate, for each model, the following quantities: the number of radio pulsars (N$_{\rm R}$), the number of \gray\ pulsars in the Parkes Multibeam Survey covered regions (N$_{\rm G}$), the number of \gray\ pulsars in the entire sky (N$_{\rm G, all-sky}$), the number of radio-loud \gray\ pulsars in the Parkes Multibeam Survey covered regions (N$_{\rm RLGL}$). These quantities are listed in columns 2 to 5 in Table~\ref{tab:number_ns}. 
As discussed before, we repeat the observation part of post-processing 10 times in order to account for stochastic variation. The average and standard deviation of these realisations are recorded for each model in Table~\ref{tab:number_ns}.

\begin{table*}
        \caption{Simulation results showing numbers of different types of pulsars, the corresponding scaled percent differences, the star formation rate and the pulsar birth rate. Columns from left to right: 
        COMPAS model;
        pulsars that are detected in radio ($N_{\rm R}$); 
        pulsars detected in \gray\ in Parkes Multibeam Survey covered region ($N_{\rm G}$); 
        pulsars detected in \gray\ in the entire sky ($N_{\rm G, all-sky}$);
        pulsars detected in both radio and \gray\ in Parkes Multibeam Survey covered region ($N_{\rm RLGL}$);
        scaling factor for the model as described in text ($F$);
        scaled percent difference between catalogue and simulated radio pulsars (SPD$_{\rm R}$); 
        scaled percent difference between catalogue and simulated \gray\ pulsars (SPD$_{\rm G}$); 
        scaled percent difference between catalogue and simulated all-sky \gray\ pulsars (SPD$_{\rm G, all-sky}$); 
        scaled percent difference between catalogue and simulated radio-loud \gray\ pulsars (SPD$_{\rm RLGL}$);
        percent difference between catalogue and simulated ratio of radio over \gray\ pulsars (PD$_{\rm Q1}$); 
        percent difference between catalogue and simulated ratio of all \gray\ pulsars over radio-loud \gray\ pulsars (PD$_{\rm Q2}$); 
        star formation rate of the model;
        inverse of pulsar birth rate (yr). 
        The first row lists the total number of pulsars in the radio and \gray\ catalogues \citep[][]{Manchester:2004bp,Fermi-LAT:2023zzt}.
        MSPs are excluded as we focus on canonical pulsars.
        Details on the calculations of percent differences are given in Section~\ref{subsec:compare_numbers}.
        For the simulations, the values and uncertainties are obtained from the average and standard deviation of 10 different realisations of each model.  }
        \label{tab:number_ns}
        \centering
        \hspace{-1cm}
	\begin{tabular}{lcccccccccccc} 
		\hline
		Model & N$_{\rm R}$ & N$_{\rm G}$  & N$_{\rm G, all-sky}$ & N$_{\rm RLGL}$ & $F$ & SPD$_{\rm G}$ & SPD$_{\rm G, all-sky}$ & SPD$_{\rm RLGL}$ & PD$_{\rm Q1}$ & PD$_{\rm Q2}$ & SFR (\Msun /yr) & 1 / PBR (yr) \\
		\hline

Catalogues & 1060 & 85 & 156 & 38 & -- & -- & -- & -- & -- & -- & -- & --  \\
INIT & $961 \pm 30$ & $37 \pm 4$ & $78 \pm 6$ & $10 \pm 2$ & $1.1$  & $ -52.2$ & $-44.6$ & $ -69.5$ & $ 109.4$ & $-36.2$ & $ 1.82$ & $ 422$  \\ 
D1 & $1051 \pm 22$ & $39 \pm 6$ & $83 \pm 7$ & $12 \pm 4$ & $1.01$  & $ -53.9$ & $-46.6$ & $ -68.1$ & $ 117.1$ & $-30.8$ & $ 1.66$ & $ 462$  \\ 
D2 & $902 \pm 40$ & $56 \pm 6$ & $122 \pm 8$ & $16 \pm 4$ & $1.18$  & $ -22.0$ & $-7.7$ & $ -50.2$ & $ 28.2$ & $-36.1$ & $ 1.94$ & $ 396$  \\ 
D3 & $959 \pm 39$ & $42 \pm 5$ & $87 \pm 7$ & $15 \pm 3$ & $1.11$  & $ -45.8$ & $-38.1$ & $ -55.2$ & $ 84.4$ & $-17.4$ & $ 1.82$ & $ 422$  \\ 
D4 & $509 \pm 21$ & $49 \pm 4$ & $92 \pm 6$ & $12 \pm 3$ & $2.08$  & $ 19.3$ & $23.2$ & $ -33.7$ & $ -16.2$ & $-44.4$ & $ 3.44$ & $ 224$  \\ 
D5 & $928 \pm 28$ & $35 \pm 3$ & $80 \pm 5$ & $11 \pm 3$ & $1.14$  & $ -52.7$ & $-41.5$ & $ -67.5$ & $ 111.4$ & $-31.4$ & $ 1.89$ & $ 408$  \\ 
B1 & $866 \pm 35$ & $43 \pm 5$ & $96 \pm 7$ & $12 \pm 2$ & $1.22$  & $ -37.5$ & $-25.0$ & $ -60.7$ & $ 60.1$ & $-37.1$ & $ 2.02$ & $ 381$  \\ 
B2 & $4501 \pm 27$ & $80 \pm 6$ & $185 \pm 4$ & $42 \pm 4$ & $0.24$  & $ -77.7$ & $-72.0$ & $ -74.2$ & $ 348.4$ & $15.9$ & $ 0.39$ & $ 1979$  \\ 
P1 & $651 \pm 21$ & $7 \pm 1$ & $20 \pm 4$ & $2 \pm 1$ & $1.63$  & $ -87.2$ & $-78.9$ & $ -89.7$ & $ 679.0$ & $-19.9$ & $ 2.69$ & $ 286$  \\ 
P2 & $965 \pm 16$ & $42 \pm 7$ & $94 \pm 5$ & $11 \pm 4$ & $1.1$  & $ -45.9$ & $-34.0$ & $ -67.3$ & $ 84.7$ & $-39.7$ & $ 1.81$ & $ 424$  \\ 
P3 & $645 \pm 25$ & $5 \pm 2$ & $24 \pm 2$ & $2 \pm 1$ & $1.64$  & $ -90.7$ & $-75.0$ & $ -93.5$ & $ 977.5$ & $-30.1$ & $ 2.71$ & $ 284$  \\ 
PN1 & $917 \pm 27$ & $37 \pm 4$ & $82 \pm 7$ & $11 \pm 2$ & $1.16$  & $ -49.8$ & $-39.1$ & $ -65.6$ & $ 99.3$ & $-31.5$ & $ 1.91$ & $ 403$  \\ 
PN2 & $948 \pm 25$ & $50 \pm 4$ & $94 \pm 7$ & $15 \pm 3$ & $1.12$  & $ -33.7$ & $-32.9$ & $ -55.9$ & $ 50.9$ & $-33.4$ & $ 1.84$ & $ 417$  \\ 
PN3 & $969 \pm 24$ & $42 \pm 6$ & $84 \pm 4$ & $13 \pm 4$ & $1.09$  & $ -45.3$ & $-41.0$ & $ -63.2$ & $ 82.9$ & $-32.6$ & $ 1.8$ & $ 426$  \\ 
PLN1 & $747 \pm 19$ & $42 \pm 3$ & $72 \pm 7$ & $13 \pm 2$ & $1.42$  & $ -30.2$ & $-34.7$ & $ -50.7$ & $ 43.3$ & $-29.4$ & $ 2.34$ & $ 328$  \\ 
BLN1 & $1042 \pm 21$ & $56 \pm 3$ & $113 \pm 8$ & $16 \pm 3$ & $1.02$  & $ -33.4$ & $-26.6$ & $ -57.2$ & $ 50.1$ & $-35.7$ & $ 1.68$ & $ 458$  \\ 
BLN2 & $2699 \pm 41$ & $70 \pm 5$ & $146 \pm 12$ & $29 \pm 5$ & $0.39$  & $ -67.6$ & $-63.3$ & $ -69.6$ & $ 208.3$ & $-6.3$ & $ 0.65$ & $ 1187$  \\ 
BLN3 & $663 \pm 22$ & $37 \pm 5$ & $94 \pm 7$ & $13 \pm 3$ & $1.6$  & $ -30.4$ & $-3.3$ & $ -43.6$ & $ 43.7$ & $-19.0$ & $ 2.64$ & $ 292$  \\ 
BLN4 & $2260 \pm 26$ & $80 \pm 4$ & $165 \pm 11$ & $31 \pm 3$ & $0.47$  & $ -55.9$ & $-50.3$ & $ -61.6$ & $ 126.8$ & $-12.9$ & $ 0.77$ & $ 994$  \\ 
P2-D1 & $1066 \pm 35$ & $52 \pm 6$ & $105 \pm 7$ & $17 \pm 3$ & $0.99$  & $ -39.2$ & $-32.8$ & $ -55.0$ & $ 64.4$ & $-26.0$ & $ 1.64$ & $ 469$  \\ 
P2-D2 & $874 \pm 18$ & $47 \pm 6$ & $121 \pm 4$ & $14 \pm 4$ & $1.21$  & $ -32.8$ & $-6.3$ & $ -54.7$ & $ 48.8$ & $-32.6$ & $ 2.0$ & $ 384$  \\ 
P2-D3 & $916 \pm 24$ & $48 \pm 6$ & $91 \pm 7$ & $14 \pm 3$ & $1.16$  & $ -35.4$ & $-32.8$ & $ -57.1$ & $ 54.7$ & $-33.6$ & $ 1.91$ & $ 403$  \\ 
P2-D4 & $521 \pm 18$ & $59 \pm 4$ & $111 \pm 6$ & $20 \pm 5$ & $2.04$  & $ 40.6$ & $45.2$ & $ 6.6$ & $ -28.9$ & $-24.2$ & $ 3.36$ & $ 229$  \\ 
P2-D5 & $959 \pm 23$ & $51 \pm 4$ & $98 \pm 7$ & $16 \pm 2$ & $1.1$  & $ -33.6$ & $-30.5$ & $ -54.3$ & $ 50.5$ & $-31.3$ & $ 1.82$ & $ 422$  \\ 
PN1-B1 & $860 \pm 36$ & $38 \pm 5$ & $76 \pm 7$ & $12 \pm 4$ & $1.23$  & $ -45.4$ & $-40.2$ & $ -60.1$ & $ 83.0$ & $-27.0$ & $ 2.03$ & $ 378$  \\ 
PN1-B1-D1 & $921 \pm 30$ & $38 \pm 4$ & $76 \pm 5$ & $13 \pm 2$ & $1.15$  & $ -48.7$ & $-44.1$ & $ -60.9$ & $ 94.8$ & $-23.9$ & $ 1.9$ & $ 405$  \\ 
PN1-B1-D2 & $824 \pm 17$ & $49 \pm 7$ & $95 \pm 5$ & $15 \pm 4$ & $1.29$  & $ -26.4$ & $-21.9$ & $ -50.6$ & $ 35.9$ & $-32.8$ & $ 2.12$ & $ 362$  \\ 
PN1-B1-D3 & $858 \pm 19$ & $44 \pm 4$ & $90 \pm 6$ & $11 \pm 3$ & $1.24$  & $ -36.3$ & $-28.5$ & $ -63.9$ & $ 57.0$ & $-43.3$ & $ 2.04$ & $ 377$  \\ 
PN1-B1-D4 & $509 \pm 20$ & $34 \pm 5$ & $80 \pm 5$ & $12 \pm 4$ & $2.08$  & $ -16.7$ & $6.9$ & $ -33.7$ & $ 20.1$ & $-20.4$ & $ 3.43$ & $ 224$  \\ 
PN1-B1-D5 & $837 \pm 28$ & $32 \pm 4$ & $73 \pm 7$ & $13 \pm 3$ & $1.27$  & $ -51.9$ & $-40.7$ & $ -55.7$ & $ 107.7$ & $-7.9$ & $ 2.09$ & $ 368$  \\ 
PN5-B1-D1 & $779 \pm 22$ & $6 \pm 1$ & $24 \pm 3$ & $3 \pm 1$ & $1.36$  & $ -90.9$ & $-79.4$ & $ -90.0$ & $ 995.5$ & $9.9$ & $ 2.25$ & $ 342$  \\ 
PN5-B1-D2 & $669 \pm 23$ & $10 \pm 3$ & $25 \pm 5$ & $4 \pm 2$ & $1.58$  & $ -82.1$ & $-74.3$ & $ -81.7$ & $ 458.9$ & $2.5$ & $ 2.61$ & $ 294$  \\ 
PN5-B1-D3 & $750 \pm 23$ & $9 \pm 2$ & $28 \pm 3$ & $4 \pm 1$ & $1.41$  & $ -84.4$ & $-75.0$ & $ -87.0$ & $ 540.1$ & $-16.7$ & $ 2.33$ & $ 330$  \\ 
PN5-B1-D4 & $366 \pm 16$ & $8 \pm 1$ & $24 \pm 3$ & $4 \pm 1$ & $2.9$  & $ -74.4$ & $-56.3$ & $ -71.8$ & $ 290.9$ & $10.4$ & $ 4.78$ & $ 161$  \\ 
PN5-B1-D5 & $757 \pm 29$ & $10 \pm 2$ & $29 \pm 5$ & $4 \pm 1$ & $1.4$  & $ -84.2$ & $-74.2$ & $ -84.5$ & $ 532.6$ & $-2.1$ & $ 2.31$ & $ 333$  \\ 
BLN1-PN1 & $1016 \pm 34$ & $37 \pm 3$ & $79 \pm 3$ & $13 \pm 3$ & $1.04$  & $ -54.2$ & $-46.9$ & $ -64.6$ & $ 118.3$ & $-22.6$ & $ 1.72$ & $ 446$  \\ 
BLN4-PN1 & $2260 \pm 35$ & $45 \pm 6$ & $121 \pm 8$ & $24 \pm 4$ & $0.47$  & $ -75.2$ & $-63.6$ & $ -69.8$ & $ 303.6$ & $22.1$ & $ 0.77$ & $ 993$  \\ 
BLN1-PN5 & $794 \pm 25$ & $11 \pm 3$ & $30 \pm 4$ & $6 \pm 3$ & $1.34$  & $ -82.6$ & $-74.7$ & $ -79.6$ & $ 473.3$ & $16.9$ & $ 2.2$ & $ 349$  \\ 
BLN5-PLN1-D4 & $234 \pm 14$ & $20 \pm 2$ & $34 \pm 3$ & $6 \pm 2$ & $4.54$  & $ 7.3$ & $0.4$ & $ -28.3$ & $ -6.8$ & $-33.2$ & $ 7.49$ & $ 103$  \\ 
BLN6-PLN2-D4 & $207 \pm 13$ & $13 \pm 2$ & $29 \pm 3$ & $3 \pm 1$ & $5.12$  & $ -21.6$ & $-5.4$ & $ -55.5$ & $ 27.6$ & $-43.2$ & $ 8.45$ & $ 91$  \\ 
		\hline
	\end{tabular}
\end{table*}

For many of the models, the predicted number of pulsars differs significantly from the observations.
For example, model B1 produces 18.3\% fewer radio pulsars, and 49.4\% fewer \gray\ pulsars, respectively.
A more extreme example is model B2, predicting 4.2 times as many radio pulsars as observed.
Due to these large differences, we do not calculate the percent difference with the numbers reported in columns 2 to 5 in Table~\ref{tab:number_ns}.
Instead, we calculate a scaled percent difference (SPD) using scaled numbers of all types of pulsars. The scaling factor is defined as $F = 1060/{\rm N_{\rm R}}$, with 1060 is the number of observed radio pulsars from the Parkes Multi Beam survey, as listed in the sixth column in Table~\ref{tab:number_ns} for each model.
These scaled numbers of \gray\ pulsar in Parkes covered regions, \gray\ pulsars observed in the entire sky, and radio-loud \gray\ pulsars are then compared to those from the catalogues to calculate the SPD. Here we give the equations on how these SPDs are calculated:
\begin{equation*}
    {\rm SPD_{\rm R}} = 100 \times \frac{{\rm N_{\rm R}} \times F - {\rm N_{\rm R,0}}}{{\rm N_{\rm R,0}}}, 
\end{equation*}
where ${\rm N_{\rm R,0}} = 1060$ is the number of radio pulsars in Parkes covered regions in the catalogue;
\begin{equation*}
   {\rm SPD_{\rm G}} = 100 \times \frac{{\rm N_{\rm G}} \times F - {\rm N_{\rm G,0}}}{\rm N_{\rm G,0}}, 
\end{equation*}
where ${\rm N_{\rm G,0}} = 85$ is the number of \gray\ pulsars in Parkes covered regions in the catalogue;
\begin{equation*}
     {\rm SPD_{\rm G,all-sky}} = 100\times \frac{{\rm N_{\rm G,all-sky} \times F - {\rm N_{\rm G,all-sky,0}}}}{\rm N_{\rm G,all-sky,0}}, 
\end{equation*}
where ${\rm N_{\rm G,0}} = 156$ is the number of \gray\ pulsars in the entire sky in the catalogue;
\begin{equation*}
     {\rm SPD_{\rm RLGL}} = 100\times \frac{{\rm N_{\rm RLGL}} \times F - {\rm N_{\rm RLGL,0}}} {\rm N_{\rm RLGL,0}}, 
\end{equation*}
where ${\rm N_{\rm RLGL,0}} = 38$ is the number of radio-loud \gray\ pulsars in the Parkes covered region in the catalogue.
The SPD values are tabulated in columns 7 to 10 in Table~\ref{tab:number_ns}.
By design, all models would have a SPD$_{\rm R}$ = 0 and their SPD$_{\rm R}$'s are hence not listed in Table~\ref{tab:number_ns}.
This scaling practice is equivalent of setting the SFR as a free parameter. As a result, the SFR for each model will be changed to $F \times 1.65$ \Msun\ yr$^{-1}$ and the PBR is changed to $F \times 2.15\times10^{-3}$ yr$^{-1}$, or 1 in every $446/F$ yr, which are listed in the second last and last columns in Table~\ref{tab:number_ns} respectively.

Another indicator of the goodness of the models is the ratio between different types of pulsars. 
We check the ratio between radio and \gray\ pulsars predicted by the models in Parkes covered region, ${\rm Q_1} = {\rm N_{\rm R}}/{\rm N_{\rm G}}$; and the ratio between all \gray\ pulsars and radio-loud \gray\ pulsars predicted by the models in the Parkes covered region, ${\rm Q_2} = {\rm N_{\rm G}}/{\rm N_{\rm RLGL}}$. Percent differences for ${\rm Q_1}$ and ${\rm Q_2}$ when compared to the observed values respectively are calculated and present in columns 11 and 12 in Table~\ref{tab:number_ns}.

A striking observation is that the PN5 series models, including PN5-B1-D1--5 and BLN1-PN5 significantly underproduce \gray\ pulsars. This indicates that the initial spin periods of \gray\ pulsars should not be as long as described in PN5 models.  

Another prominent feature is that all models, except D4 and P2-D4, underproduce \gray\ pulsars to a certain degree. 
It could be because of the selection criteria for \gray\ pulsars is too stringent, and this is discussed in \S~\ref{subsec:comparison}. A strange phenomenon is that the degree of underproduction of \gray\ pulsars decreases when comparing all-sky \gray\ pulsars, instead of Parkes covered \gray\ pulsars.  

Based on the percent differences across the various pulsar types, we consider models with all percent differences below 60\% to be satisfactory. With this criterion we identify the following models to be reasonable: D2, D4, PN2, PLN1, BLN1, BLN3, P2-D2, P2-D3, P2-D4, P2-D5, PN1-B1-D2, PN1-B1-D4, BLN5-PLN2-D4 and BLN6-PLN1-D4. Among these, PN1-B1-D4 and BLN5-PLN2-D4 exhibit the lowest SPDs across all categories of pulsar types, followed by D2, P2-D4, PN1-B1-D2, and BLN6-PLN3-D4.
These results indicate that models with lower $\tau_d$ produces the right number of pulsars. However, some of the other models with larger $\tau_d$ also perform reasonably well in this regard. The other models listed above with reasonable numbers of different types of pulsars suggest that models with high birth magnetic field and short birth spin periods are more likely to produce the right numbers of different types of pulsars. 

\subsection{Best Fit Model}
\label{subsec:bestfit}

Combining the results presented in Tables~\ref{tab:pvalues-radio}, ~\ref{tab:pvalues_gamma} and ~\ref{tab:number_ns}, we identify models that best describe both the radio and \gray\ pulsar populations.
Ideal models need to explain the observed distributions of the physical properties and reproduce the numbers of different pulsar types.
However, no single model satisfies all three criteria perfectly.
Among the models, PN1-B1-D4 is the best fit, as it is listed as one of the top models for both the radio and gamma-ray populations.
This model produces the numbers of different pulsar types and their ratios that are comparable to the observations.
The star formation rate is 3.43 \Msun\ yr s$^{-1}$, agreeing with the values listed in \S~\ref{subsec:sfrate}. The pulsar birth rate of 1 per 224 yr, while low compared to the literature values, is comparable to the rate of CCSN in COMPAS models, as described in \S~\ref{subsec:sfrate}. A degeneracy between the number of pulsars formed in the model and the fraction of observable ones is expected. With higher pulsar birth rate, a smaller fraction is needed to produce the same amount of observable pulsars.
This effect will be explored in the future once the issue of the low CCSN rate has been addressed in COMPAS.
Model BLN5-PLN2-D4 is a strong contender for the best fit model, describing the distributions of physical parameters of both radio and \gray\ pulsars well, and having the lowest percent differences in almost all types of pulsars and the two ratios. However, the SFR at 7.49 \Msun\ yr s$^{-1}$ is considerably high compared to the literature values, even though the resulting pulsar birth rate at 1 per 103 yr is much closer to the literature values as listed in \S~\ref{subsec:sfrate}. This rules it out for consideration of the best fit model.

\begin{figure*}
    \hspace*{-2.5cm}
    \includegraphics[width=1.3\textwidth]{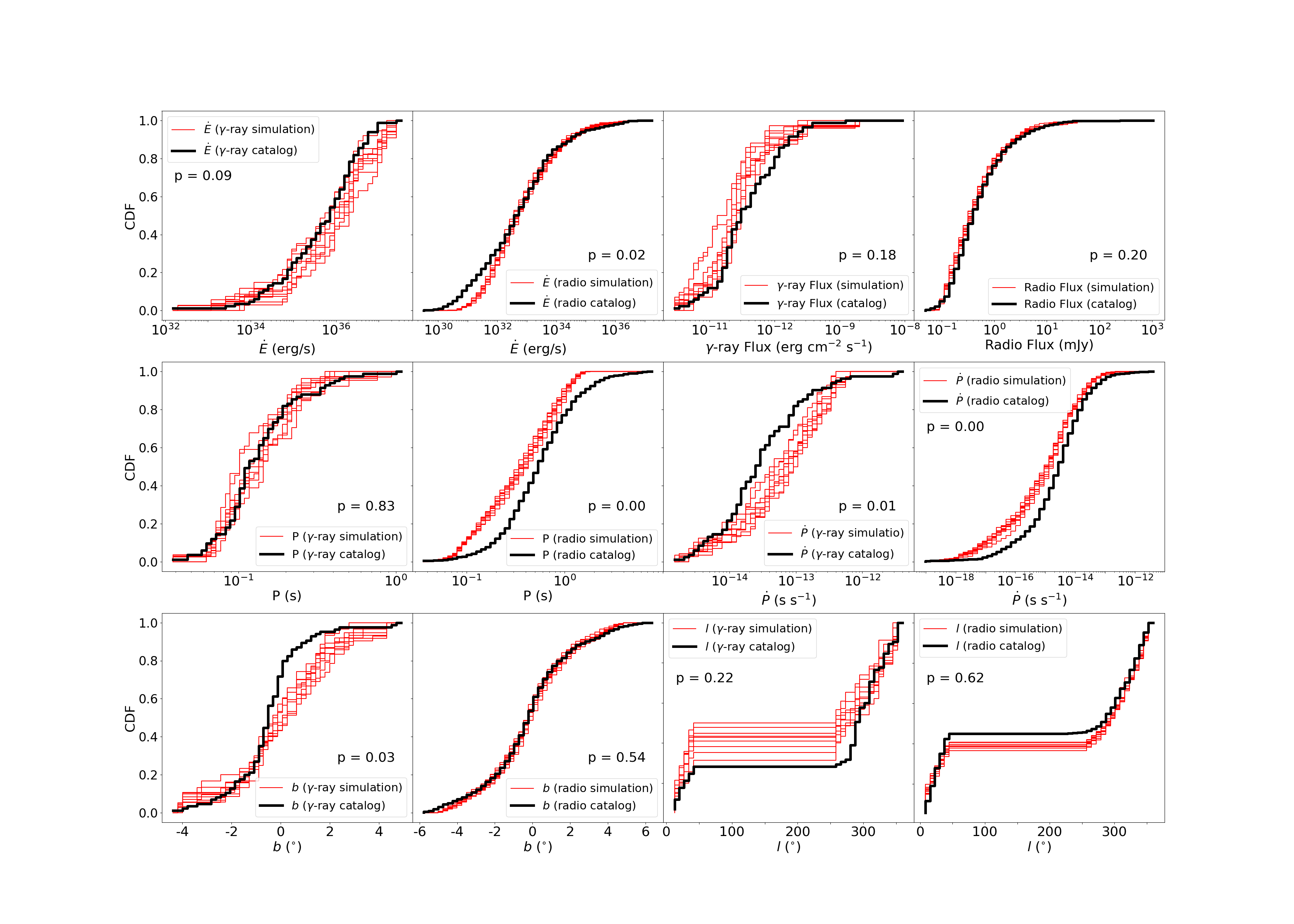}
    \caption{CDFs of various physical quantities predicted/produced by the PN1-B1-D4 model (red), compared to the data from the catalogues (blue). The red lines in these plots are from all ten realisations of the PN1-B1-D4 model that are sampled randomly as described in \S~\ref{subsec:reusing_COMPAS}.
    From top left to bottom right, the panels show the distributions of  \gray\ pulsar \Edot , radio pulsar \Edot  ,  \gray\ flux, radio flux, \gray\ pulsar period, radio pulsar period, \gray\ pulsar \Pdot, radio pulsar \Pdot, \gray\ pulsar galactic latitude, radio pulsar galactic latitude, \gray\ pulsar galactic longitude and radio pulsar galactic longitude. The p-values for each individual parameter as listed in Tables~\ref{tab:pvalues-radio} \& ~\ref{tab:pvalues_gamma} for one random realisation are plotted on each respective panel.}
    
    \label{fig:pn1b1d4_results}
\end{figure*}

To further examine model PN1-B1-D4, we plot the CDFs of the quantities listed in Tables.~\ref{tab:pvalues-radio} and~\ref{tab:pvalues_gamma} in Figure~\ref{fig:pn1b1d4_results}, showing the results from all 10 realisations. 
The CDFs of some of the parameters, such as spin period of \gray\ pulsars, are well matched to the observations, while some others such as spin period of radio pulsars, are not. This is expected when examining Tables~\ref{tab:pvalues-radio} \&~\ref{tab:pvalues_gamma}.

\begin{figure}
    \centering
    \includegraphics[width=\columnwidth]{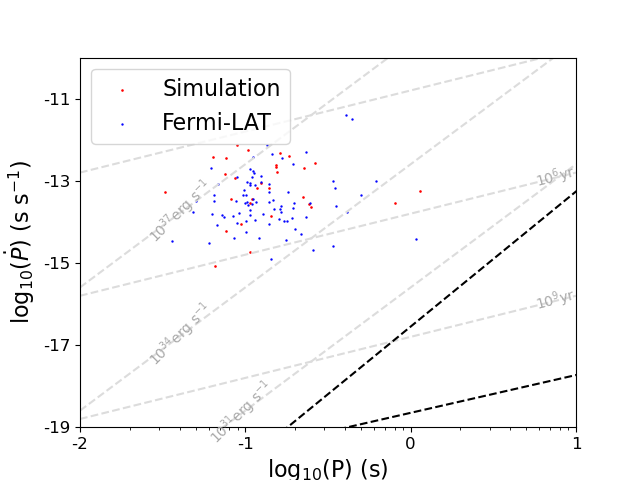}
    \includegraphics[width=\columnwidth]{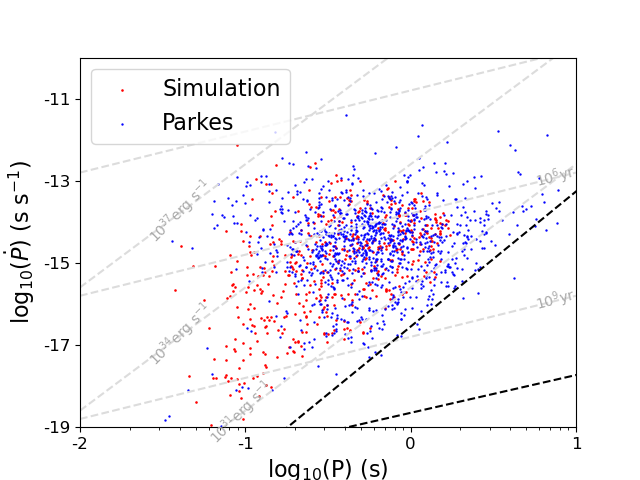}
    
    \caption{Top: P-\Pdot\ diagram of {\it Fermi} detected canonical pulsars (blue) and detected \gray\ pulsars from the characteristic PN1-B1-D4 model (red); bottom: P-\Pdot\ diagram of radio detected canonical pulsars as recorded in the ATNF catalogue (blue) and detected radio pulsars from the characteristic PN1-B1-D4 model (red). The two black dashed lines represent the death-lines described in Eqs.~\ref{eq:dl1} \& ~\ref{eq:dl2}. The gray dashed lines labelled $10^3$ yr, $10^6$ yr, and $10^9$ yr represent constant characteristic age each respectively; and those labelled with $10^{31}$ \LumUnit\ , $10^{33}$ \LumUnit\ , and $10^{37}$ \LumUnit\ represent constant \Edot\ each respectively. }
    \label{fig:ppdot-two}
\end{figure}

Additionally, we visualise the results with the P-\Pdot\ diagrams for \gray\ and radio pulsars generated from one randomly chosen realisation of the PN1-B1-D4 model, overlaid with the canonical pulsars from the catalogue, as shown in Figure~\ref{fig:ppdot-two}. 
These diagrams provide a clearer comparison between the modelled and observed populations.
For \gray\ pulsars, the simulation and catalogue pulsar populations occupy the same part of the parameter space.
The simulated radio pulsars, compared to the observation, are shifted towards the bottom left of the P-\Pdot\ diagram. This suggests that the simulated radio pulsars have lower magnetic fields compared to observation. This is anticipated given the PN1-B1-D4 model has a small $\tau_d$, letting the magnetic field to decay faster, resulting in pulsars with lower field and shorter spin period.

\begin{figure*}
    \centering
    \includegraphics[width=0.9\textwidth]{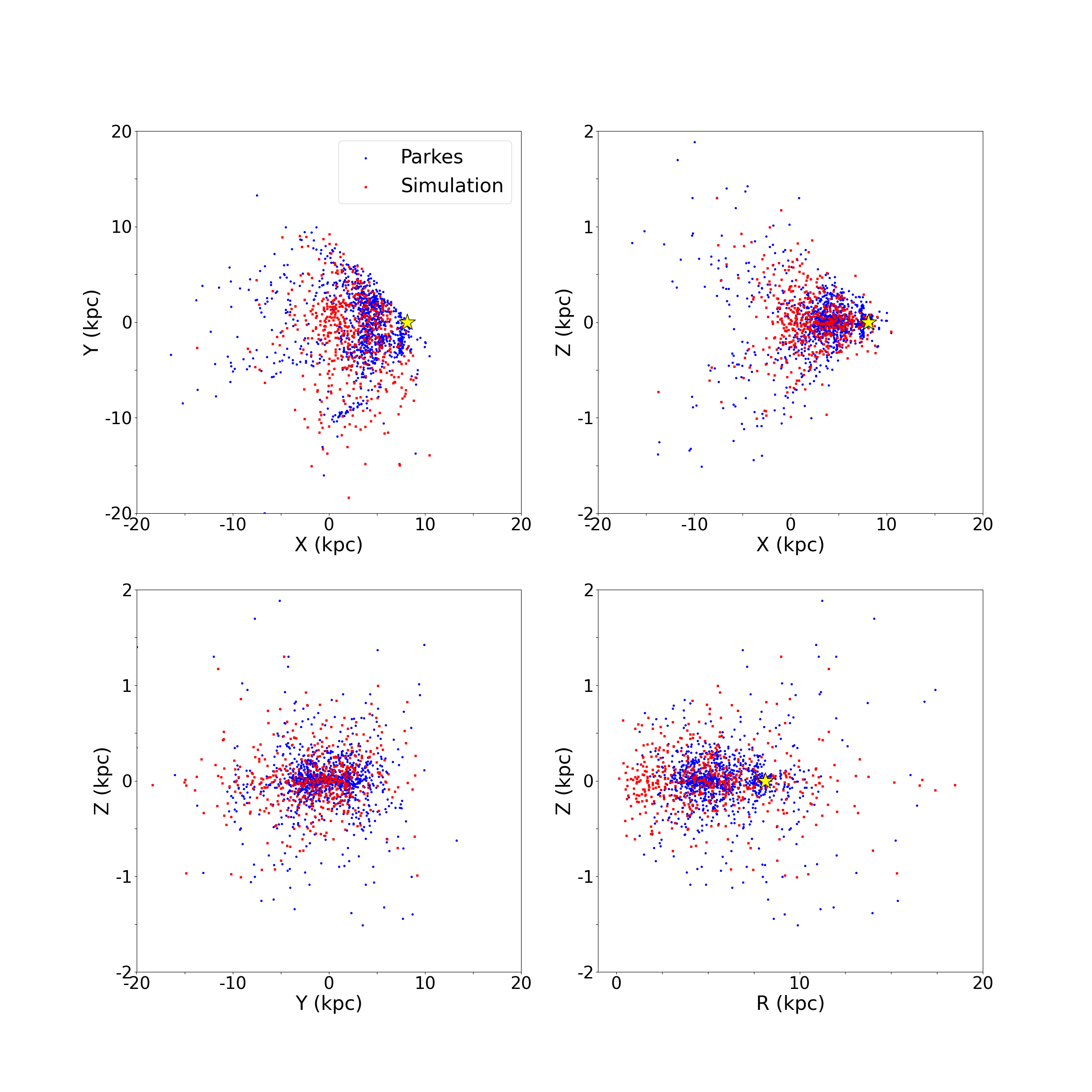}\\
    \caption{Galactocentric coordinates of radio pulsars from the catalogue (blue) and from PN1-B1-D4 model (red). The yellow star in all panels except for the Y-Z planes indicates the location of the Sun.}
    \label{fig:locs_r}
\end{figure*}

\begin{figure*}
    \includegraphics[width=0.9\textwidth]{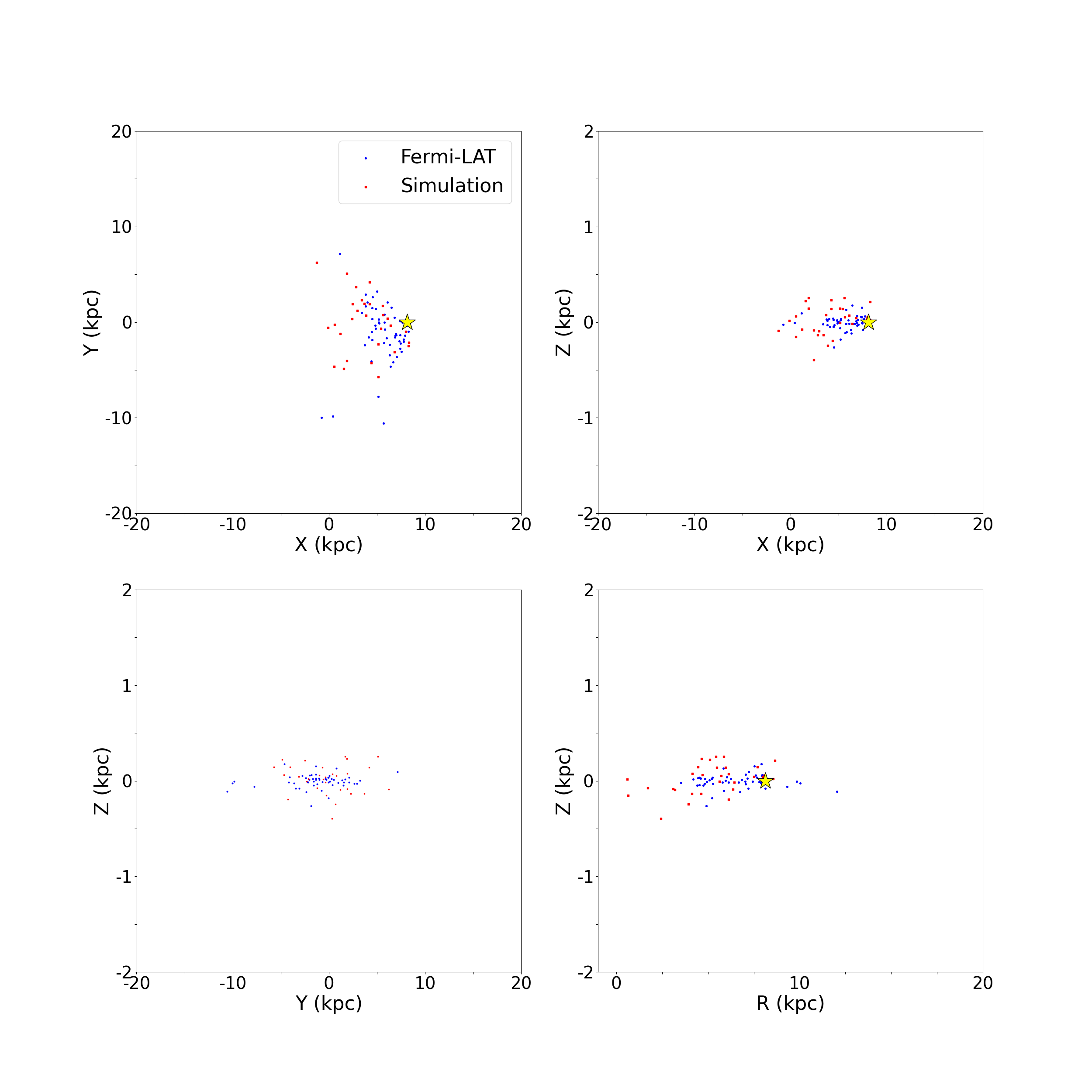}\\
    \caption{Galactocentric coordinates of \gray\ pulsars from the catalogue (blue) and from PN1-B1-D4 model (red). The yellow star in all panels except for the Y-Z planes indicates the location of the Sun. }
    \label{fig:locs_g}
\end{figure*}

We show the spatial coordinates of detected pulsars in Figures~\ref{fig:locs_r} \& ~\ref{fig:locs_g}. In both Figures~\ref{fig:locs_r} \& ~\ref{fig:locs_g}, the spatial distributions of pulsars in the X-Y, X-Z and Y-X planes have good overlap between the observation and the simulation, with a slight over density around the Galactic Centre region. This over density is the most obvious in the R-Z plot at the bottom right panel in both Figures~\ref{fig:locs_r} \& ~\ref{fig:locs_g}. We attribute this to the initial spatial distributions of massive stars, but overall the Galactic model used in this work is sufficiently good.


\subsection{Magnetic Decay Timescale}
\label{subsec:taud}

Magnetic decay timescale ($\tau_d$) is a key parameter for pulsar evolution models. 
As mentioned in \S~\ref{subsec:free_params}, there is no consensus on the value of $\tau_d$. 
Some studies, mentioned in \S~\ref{subsec:free_params}, suggest short $\tau_d$, indicating rapid magnetic field decay; while others prefer long $\tau_d$, indicating a slow-changing or even constant magnetic field throughout the life time of a NS. 
In this section, we use population synthesis methods to examine both perspectives. 
As shown in Table~\ref{tab:compas_parameter}, four sets of models are examined: INIT, P2, PN1-B1 and PN5-B5, each with specific birth magnetic field and spin period distributions and $\tau_d$ ranging from 1 Myr to 10,000 Myr.

For radio pulsars, shorter $\tau_d$ values (1--10 Myr) generally produce better matches to observations across most model sets, while longer $\tau_d$ values (100--10,000 Myr) tend to yield worse fits.
The caveat of having shorter $\tau_d$ for radio pulsars is that the models tend to generate more low field pulsars, resulting with a population with shorter spin periods.
For \gray\ pulsars, moderate to long $\tau_d$ values (e.g., > 100 Myr) better match the observed population, as slow-decaying or effectively non-decaying magnetic fields are more consistent with the high \Edot\ s and low characteristic ages of these pulsars.
Notably, model set PN1-B1 stands out for its strong performance in describing radio pulsar populations, with $\tau_d = 1$\,Myr emerging as the best fit. 

Overall, the results favour shorter $\tau_d$ values (1--10 Myr) for describing the population of neutron stars, particularly radio pulsars, as these models consistently align with observed distributions of pulsar properties, as well as the numbers and ratios of different pulsar types. However, \gray\ pulsars require longer $\tau_d$ values for optimal fits. 

One caveat is that this result is based on the assumption that the magnetic field always decays exponentially with time, as described in \S~\ref{subsec:pulsar_evolution}. 
The exponentially decaying field in our model suggests that different decay timescales are needed to explain older radio and younger gamma-ray pulsars. However, accounting for the Hall effect at earlier times, as mentioned in \S~\ref{subsec:pulsar_evolution}, would bridge this issue. 
This would require modelling the magnetic field decay with different prescriptions \citep[e.g.,][]{Colpi:2000ApJL,DallOsso:2012MNRAS} and it is planned for future work (see \S~\ref{sec:future}).

\subsection{Impact of \gray\ Beaming and Sensitivity}
\label{subsec:comparison}

Two model choices can impact the simulation results: \gray\ beaming and the \gray\ sensitivity limit.

\begin{figure}
    \centering
    \includegraphics[width=\linewidth]{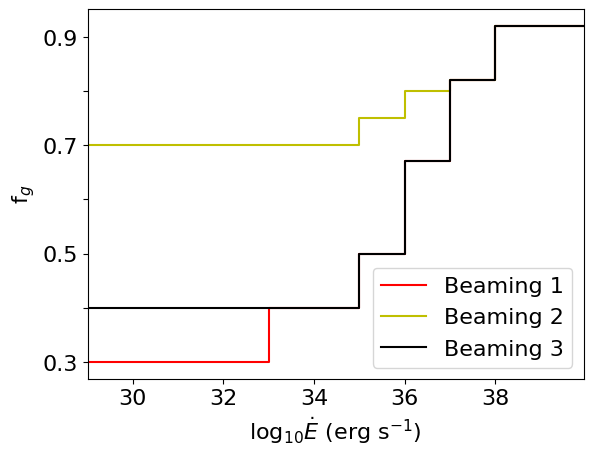}
    \caption{Three models for the beaming of \gray\ pulsars as a function of \Edot\ as described in Section~\ref{subsec:comparison}.}
    \label{fig:beaming}
\end{figure}

Our initial suspicion for the underproduction of \gray\ pulsars was that it was due to the beaming factor (as described in \S~\ref{subsubsec:gamma_ray_beaming}). 
Since \gray\ beaming is applied as a probabilistic rejection scheme, variations in the beaming will impact the numbers of \gray\ pulsars. 
We made modifications to the beaming prescription outlined as follows: 
Beaming 1 option is the one described in \S~\ref{subsubsec:gamma_ray_beaming}.
Beaming 2 assumes the beaming fraction remains high to a floor of f$_g$ = 0.7 for all pulsars with \Edot\ < $10^{35}$ \LumUnit . 
For pulsars with $10^{35}$ \LumUnit\ < \Edot\ < $10^{36}$ \LumUnit , f$_g$ = 0.75. For pulsars with $10^{36}$ \LumUnit\ < \Edot\ < $10^{37}$ \LumUnit , f$_g$ = 0.8. 
Beaming 3 decreases the beaming fraction of pulsars by steps in the same way as Beaming 1 until a floor at f$_g$ = 0.4 at \Edot\ = 10$^{35}$ \LumUnit . These beaming models are simply ad hoc generalisation made to compare different prescriptions. All three beaming options are plotted in Figure~\ref{fig:beaming}.

\begin{figure*}
    \centering
    \includegraphics[width=\linewidth]{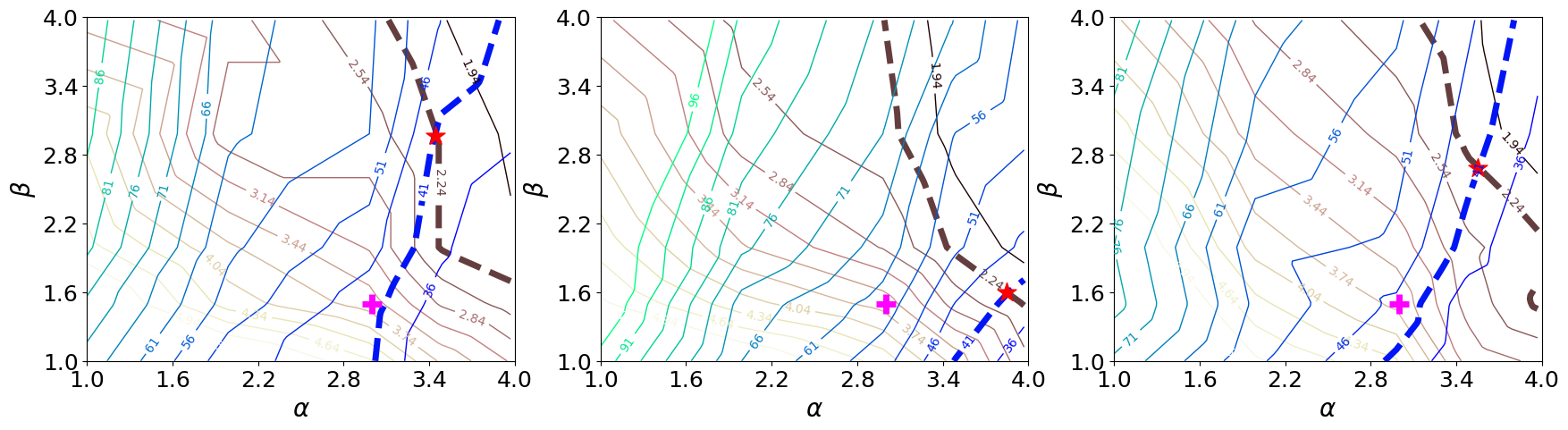}
    \caption{Results showing N$_{G}$ (yellow-brown contours) and q (blue-green contours) with different combination of $\alpha$ and $\beta$, under Beaming 1 (left), Beaming 2 (middle) and Beaming 3 (right) options, respectively. The catalogue values of N$_G$ = 41 and q = 2.24 are labelled as dashed lines in the contours. The fuchsia cross in each panel represents the default choices of $\alpha = 3.0$ and $\beta = 1.5$ used in the analysis as stated in \S~\ref{subsubsec:gamma_ray_sensitivity}. The red star in each panel denotes the intersection where the simulation results match both N$_G$ and q.  }
    \label{fig:sens_and_beam}
\end{figure*}

The sensitivity limit for radio-detected \gray\ pulsars, as described in \S~\ref{subsubsec:gamma_ray_sensitivity}, can also affect the final results. 
Specifically, when we change our assumption about the sensitivity limit, the detected population in the model will also change. 
As stated in \S~\ref{subsubsec:gamma_ray_sensitivity}, we deviate from the standard 4FGL sensitivity based on the radio detectability of the \gray\ pulsar. Here we explore the effects of varying $1.0 \leq \alpha \leq 4.0$ and $1.0 \leq \beta \leq 4.0$, then re-run the analysis pipelines with all three beaming options on the PN1-B1-D4 model. We compare the resulting number of \gray\ pulsars produced in each instance (N$_{G}$) and the ratio of \gray\ pulsars and radio-loud \gray\ pulsars ($q$) to the catalogue values in Figure~\ref{fig:sens_and_beam}. 

The results of this analysis are plotted in Figure~\ref{fig:sens_and_beam} with contours of N$_G$ and q for different combinations of $\alpha$ and $\beta$ for all three beaming options. Assuming the catalogue contains the same number of radio pulsars as predicted by the PN1-B1-D4 model (509; see Table~\ref{tab:number_ns}), the number of catalogue \gray\ pulsars is scaled down proportionally from 85 to 41, reflecting the same reduction ratio as applied to the radio pulsars (from 1060 to 509), to ensure a fair comparison.
N$_G$ = 41, and the catalogue value of q = 2.24 are both plotted with dashed lines in the contours. 
Firstly, the impact of beaming is examined. 
Beaming 2 increases number of \gray\ pulsars for all combinations of $\alpha$ and $\beta$, whilst showing a slight decrease in q in most cases. Beaming 3 shows marginal difference from Beaming 1.

For a given beaming option, when $\alpha$ increases, fewer \gray\ pulsars are detected, and q decreases. 
In this case, the decrease of N$_G$ can be attributed to radio-quiet pulsars being harder to detect in \gray s without existing timing solutions. When $\beta$ increases, more \gray\ pulsars are detected, but q decreases. 
This can be attributed to more radio pulsars being detected in \gray s. 
In all three different beaming options, high $\alpha \gtrsim 2.8$ is required to achieve the catalogue values of q and N$_G$. This agrees with the initial assumption that according to the 2PC/3PC catalogues \citep{Fermi-LAT:2013svs, Fermi-LAT:2023zzt}, {\it Fermi}-LAT is less sensitive to pulsars discovered without known radio counterparts (e.g. through blind searches). In beaming options 1 and 3, high $\beta$ is required to achieve the catalogue values, which suggests that pulsars with prior radio detection is much easier to be detected in {\it Fermi}-LAT. In beaming option 2, while only a low $\beta$ is required, the beaming fractions for pulsars with different \Edot\ are all raised significantly. This is another indication that radio pulsars should be easier to detect in \gray\ to agree with the catalogues. In comparison, if choosing the unaltered {\it Fermi}-LAT sensitivity with $\alpha = \beta = 1$, the models overproduce \gray\ pulsars, and also produce more radio quiet pulsars indicated by higher q.

Future development in better physical understanding of \gray\ beaming is required. It is also important to establish a correlation between the radio and \gray\ detection limit from the current population, which can further improve results for future population study.

\subsection{Low $\dot{E}$ Pulsars in \gray s}
\label{subsec:low_edot}

Theoretical studies of \gray\ emission from pulsars, such as the equatorial current sheet model \citep{current_sheet0, current_sheet1, current_sheet2} or the outer gap model \citep{outer_cheng1986a, outer_cheng1986b, outer_romani1996, Takata2011}, seem to indicate that pulsars with \Edot\ $< 10^{33}$ erg/s do not emit in \gray\ s. 
Indeed, if we investigate Eq.~\ref{eq:L_gamma}, it shows that at \Edot\ $= 10^{33}$ erg/s, the \gray\ luminosity of a pulsar is equal to \Edot\ , reaching maximum efficiency. 
With \Edot\ $< 10^{33}$ \LumUnit\ , if we still follow the fundamental plane relation, the \gray\ luminosity of the pulsar becomes larger than \Edot\ , breaking conservation of energy. 
It is also worth noting that according to \citet{Pulsar_FP2019, Pulsar_FP2022}, the fundamental plane relations are rooted in the equatorial current sheets model with theoretical foundations.

However, the most up-to-date 4FGL/3PC catalogues include a handful of pulsars that are detected with \Edot\ $<10^{33}$ erg/s. 
Given the low number of detections of \gray\ pulsars with \Edot\ $<10^{33}$ erg/s, there is no clear understanding of their emission mechanism. 
In \citet{Song2023}, a stacked signal is detected from low \Edot\ pulsars at 4.8$\sigma$. 
Applying the updated stacking methods from \citet{Henry:2024MNRAS} boosts the detection significance to 7$\sigma$.
However, an emission mechanism was not determined due to large uncertainty in the results. We replicate the stacking analysis performed in \citet{Song2023} for the simulated pulsars, in an attempt to determine the possible emission mechanism for low \Edot\ pulsars. 
The \gray\ stacking analysis in \citet{Song2023} is performed on the undetected Galactic pulsars outside of globular clusters. The target pulsars are more than 20\degree\ away from the Galactic plane. No selections  were made based on pulsar type. For this part of the discussion, we also need the population synthesis models to produce pulsars over the entire sky. For this reason, we assume that the radio pulsars are discovered with a ``global'' Parkes survey that has the same sensitivity but covers both the northern and southern hemispheres. \gray\ pulsars are selected as described in \S~\ref{subsubsec:gamma_ray_sensitivity}.

We choose pulsars from the PN1-B1-D4 model that are radio-loud, \gray\ quiet, that are at least $20^{\circ}$ away from the Galactic plane for this study, consistent with \citet{Song2023}. 
There are 46 such pulsars in the randomly chosen realisation of the simulation. 
We then assign each pulsar with two different \gray\ luminosities, one that follows the fundamental plane if $\dot{E} > 10^{33}$ \LumUnit and $L = 0.8 \dot{E}$ when $\dot{E} < 10^{33}$ \LumUnit; and the other one is assigned a luminosity drawn from a log normal distribution $\log_{10} \sim N(32, 0.5)$ (ergs/s). 
The former luminosity assignment corresponds to beamed \gray\ emission correlated to the \Edot\ of each pulsar, and the latter corresponds to the universal, weak and isotropic \gray\ emission. 
For the first emission mechanism, we still account for the \gray\ beaming effect discussed in \S~\ref{subsubsec:gamma_ray_beaming} when being stacked, while the isotropic emission does not need to be beamed. 

\begin{figure*}
    \centering
    \includegraphics[width=0.32\textwidth]{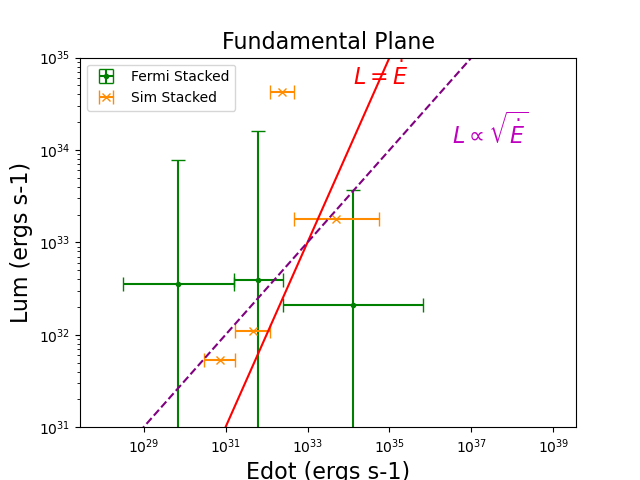}
    \includegraphics[width=0.32\textwidth]{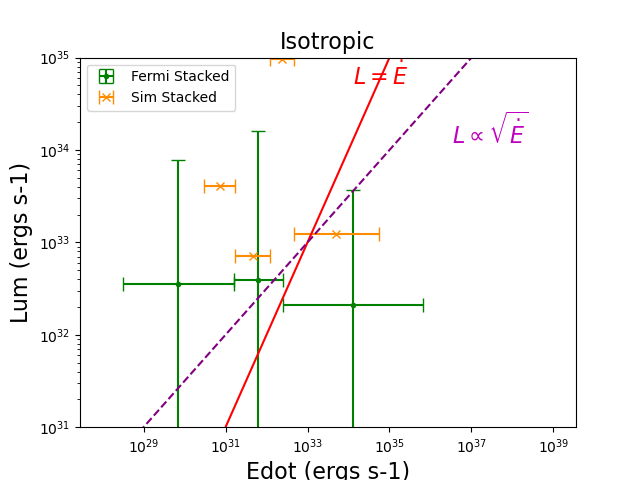}
    \includegraphics[width=0.32\textwidth]{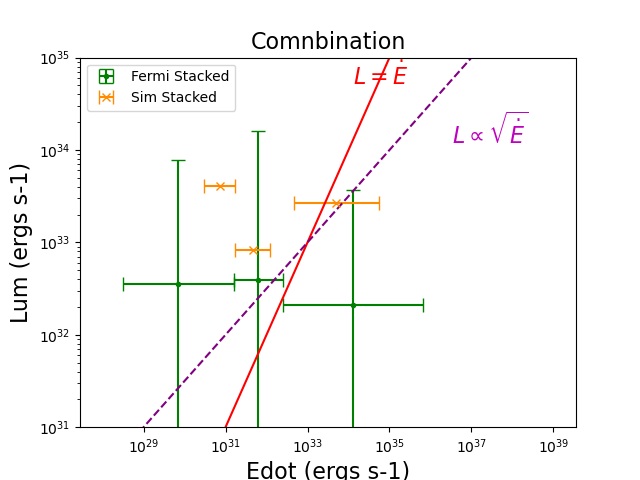}
    \caption{\gray\ luminosity vs. \Edot\ for stacked pulsars. The green data points with error bars in both directions are from \citet{Song2023} and the orange data points with only horizontal error bars are from the simulation using PN1-B1 model.  The magenta dashed line represents the heuristic relation $L_{\gamma} = \sqrt{10^{33} \dot{E}}$ and the red solid line represents the limit of energy conservation as $L_{\gamma} = \dot{E}$. The left panel calculates the stacked luminosity from the simulation by assuming the stacked pulsars have a luminosity that follows the fundamental plane while $\dot{E} > 10^{33}$ \LumUnit\ and $L_{\gamma} = 0.8 \dot{E}$ when $\dot{E} < 10^{33}$ \LumUnit\. The middle panel is calculating the luminosity assuming all these pulsars have a weak, isotropic \gray\ emission follows a log-normal distribution where $\log{L_{\gamma}} ~ N(32, 0.5)$. The right panel combines both. }
    \label{fig:stacked_lum}
\end{figure*}

We then calculate the average luminosity of the pulsars in four different \Edot\ bins, with the lowest two \Edot\ bins have the same number of pulsars and highest two \Edot\ bins having the same number of pulsars. 
We check these results in a similar way to Figure 6 in \citet{Song2023}. 
Here in this work, we choose the canonical pulsars from \citet{Song2023} and plot them in Figure~\ref{fig:stacked_lum}, and use this as the observational results to be compared with the model. 
We then plot in Figure~\ref{fig:stacked_lum} the average luminosity vs. \Edot\ from the model with three different scenarios: only the fundamental plane is considered, only isotropic emission is considered, and combining both. 
While the observational results are largely uncertain, they are consistent with all three scenarios proposed in this work.

\begin{figure}
    \centering   
    \includegraphics[width=\columnwidth]{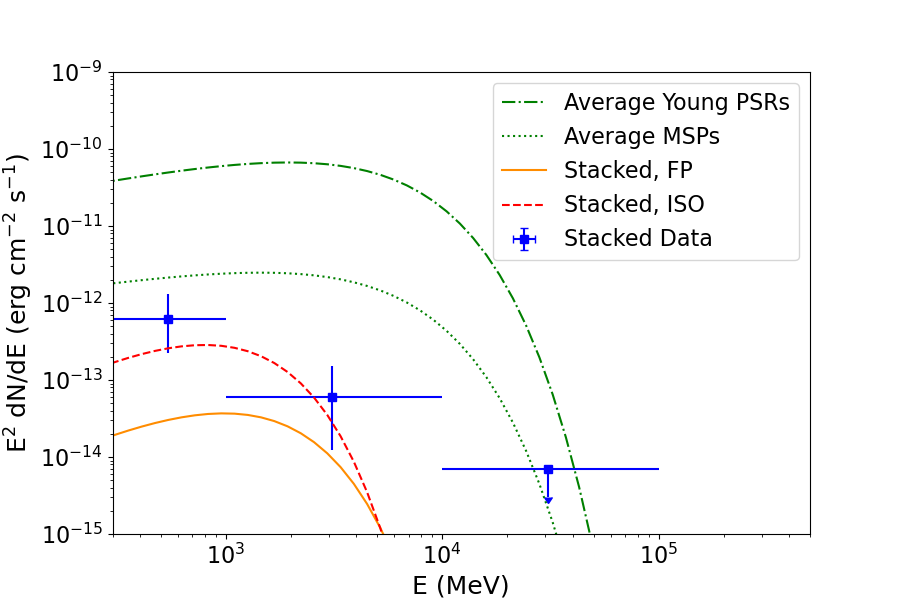}
    \caption{SED of stacked pulsars from the simulation using the PN1-B1-D4 model. The orange solid curve is the averaged SED from the stacked pulsars in the simulation assuming they all follow the empirical fundamental plane emission description in \gray s, and the red dashed curve is for the same population of pulsars assuming they all emit weakly isotropic \gray\ emission (ISO). The blue data points with error bars are the averaged SED of the sampled canonical pulsars from \citet{Song2023}, The green dotted-dashed curve represents the averaged SED of young pulsars, and the green dotted curve SED of MSPs \citep{2015ApJ...804...86M}.   }
    \label{fig:SED}
\end{figure}

We further examine the results and potential emission mechanism is to compare the spectral energy distributions (SEDs) for various scenarios. We calculate the stacked SED of the 46 pulsars for the fundamental plane emission and isotropic emission as described above. The \gray\ spectral parameters of the pulsars, $E_{\textrm{cut}}$ and PL index before cutoff, are drawn from the described text above. When considering the fundamental plane emission, a pulsar needs to be beamed towards the observer. When considering the isotropic emission, it does not need to be beamed. Averaging the SEDs, we can produce the stacked SEDs for both scenarios as shown in Figure~\ref{fig:SED}. We compare this result to \citet{Song2023} as well as the averaged pulsar SED from \citet{2015ApJ...804...86M}. 
If considering the FP emission, this stacked pulsar population is relatively older compared to the young, energetic pulsars, hence the SED is much lower. Interestingly, when considering the isotropic emission, the SED is more luminous compared to the FP emission, and roughly agrees with the stacked pulsars from the data. By examining the stacked SEDs of modelled pulsars, it seems to indicate that the isotropic emission is favoured over the fundamental plane emission. However the caveat here is that the model produces a lot fewer pulsars that satisfy the stacking criteria.

\begin{figure*}
    \centering
    \includegraphics[width=0.6\textwidth]{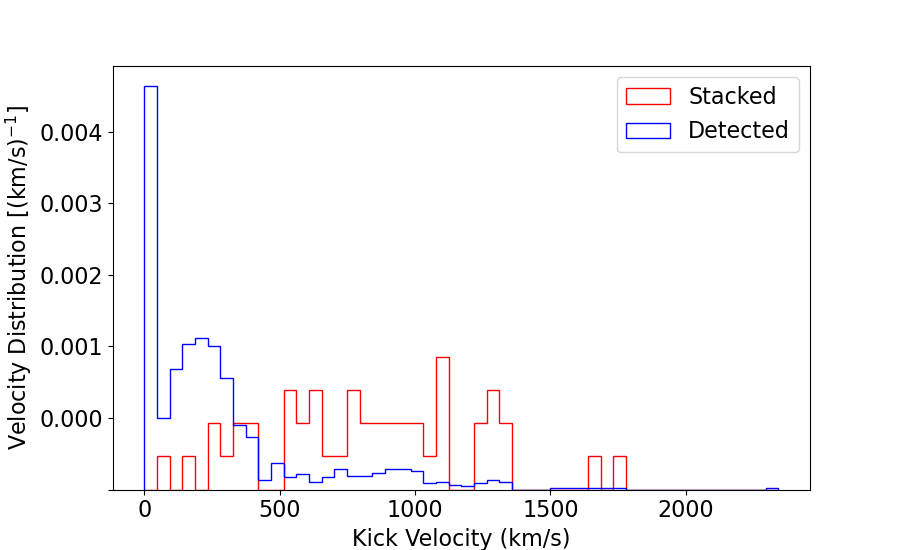}
    \caption{Kick distributions of simulated pulsars. The blue histogram shows the kick distribution of the pulsars detected in the simulation, and the red histogram shows the kick distribution of the stacked pulsars from the simulation. The selection of pulsars off of the Galactic plane preferentially selects pulsars born with large kick velocities.}
    \label{fig:kick}
\end{figure*}

Many low \Edot\ pulsars are located outside of the Galactic plane \citep{Fermi-LAT:2023zzt}. It is commonly believed that the high latitude pulsars are a result of pulsars receiving larger kicks when they went through supernova explosions. 
The results from this work can be used to confirm if that is indeed the case, as the kick velocities are one of the modelled components in our population synthesis models. 
In Figure~\ref{fig:kick}, we show the distributions of kick velocities received by pulsars from the model that are either detected or stacked, as selected above.
Consistent with our hypothesis, we find that the high latitude, low $\dot{E}$ pulsars in the stacked sample received larger natal kicks than the typical detected population.

In \citet{Smith:2019ApJ}, a discussion of distance biases on the pulsar catalogue was presented. Lower \Edot\ pulsars, the majority of which are off the Galactic plane, cannot be detected easily due to their low efficiency, small sample, and possible large distance. However, \citet{Smith:2019ApJ} suggest that if low \Edot\ pulsars emit in \gray s, it would be useful to use the population synthesis approach to help understand if a large population of such pulsars would contribute to a part of the diffuse Galactic \gray\ emission. 
Diffuse Galactic emission has been observed in GeV energies by {\it Fermi}-LAT \citep{Abdo:2009ApJ} and TeV energies \citep[e.g. by HESS;][]{Abramowski:2014PhRvD}, with its origin still under research \citep{diMauro:arXiv16} and unresolved pulsars being potential contributors \citep{Smith:2019ApJ}.
This work does not provide a full population modelling of pulsars, especially old pulsars, as only the recent star formation history is considered. Therefore we are unable to determine if the diffuse background can be attributed to old pulsars. This will be discussed in a follow-up study. 

\begin{figure}
    \includegraphics[width=\columnwidth]{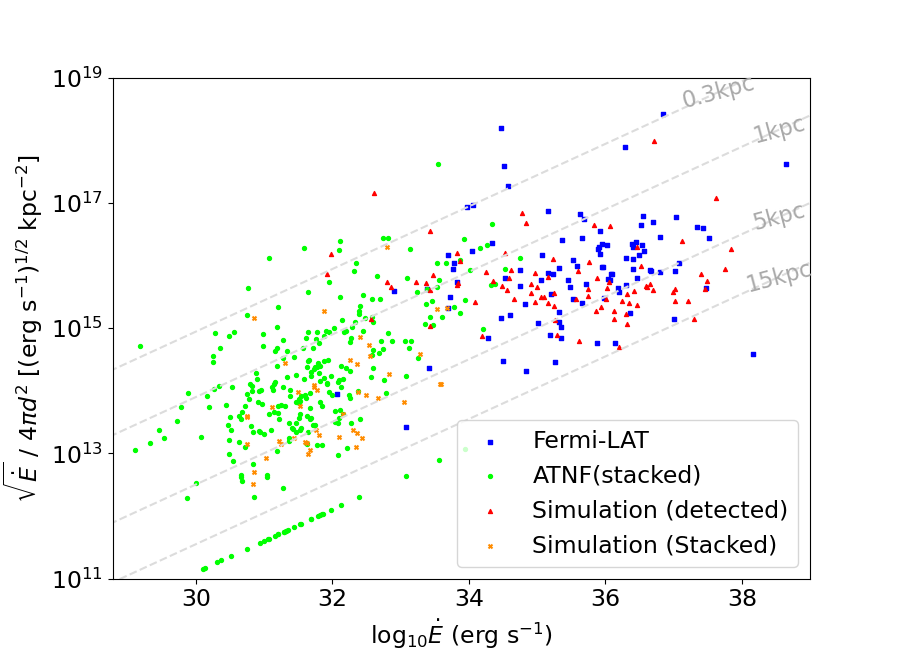}\\
    \caption{Expected \gray\ flux of pulsars (the heuristic flux) as a function of the spindown luminosity $\dot{E}$ (similar to Figure 7 in \citealp{Smith:2019ApJ}).
    We plot the expected \gray\ flux of pulsars (the heuristic flux) when assuming $L_{\gamma} = \sqrt{10^{33}\dot{E}}$ erg s$^{-1}$. 
    Blue squares are {\it Fermi}-LAT detected pulsars, red triangles are the detected pulsars from the simulation using the PN1-B1 model, green dots are the pulsars from the sample list of \citet{Song2023}, and the orange crosses are the pulsars that satisfy the criteria of being stacked from the simulation using the PN1-B1 model. The gray dotted lines labelled as 0.3 kpc, 1 kpc, 5 kpc and 15 kpc represent those of constant distance respectively. Most of the pulsars stacked in \citet{Song2023} with \Edot\ $< 10^{33}$ \LumUnit\ remain undetected in spite of having similar heuristic fluxes to those detected with higher \Edot . Most of the pulsars being stacked from the simulation have larger distances compared to those from \citet{Song2023}.}
    \label{fig:efficiency}
\end{figure}

Here we show the heuristic \gray\ flux of detected and stacked pulsars in Figure~\ref{fig:efficiency}, as for the detected pulsars, both those from the simulation and the {\it Fermi}-LAT catalogue occupy similar parts of the parameter space. 
The heuristic \gray\ flux is defined as $\sqrt{\dot{E}}/4\pi d^2$, and is often used in observation to estimate the potential \gray\ flux of a pulsar, especially if it is not detectable. 
We exclude pulsars that are estimated to have distances equal to  25\,kpc, where the distances are estimated based on the dispersion measure of radio observations, and rely on an underlying electron density model \citep{YMW17}.
In Figure~\ref{fig:efficiency}, the stacked pulsars from the simulation, in general, occupy similar ranges of distances compared to those from the catalogue. However the distribution of the distances from the stacked pulsars in the simulation peaks at larger distance compared to those from the catalogue.  

\section{Future Development}
\label{sec:future}

As the first of a series of studies using COMPAS to study \gray\ pulsar populations, there are some caveats as discussed above, and these are summarised below: 

\begin{enumerate}
    \item This work currently does not include MSP populations. Pulsar recycling physics will be implemented in a direct follow-up of this work so that a full population study of pulsars including MSPs can be carried out. MSPs are products of binary evolution, and so comparing to binary evolution models will be important to fully understand the population. 
    
    \item Some NSs are formed through accretion-induced collapses, and these NSs are not accounted for in this work. 
    We exclude them as neutron stars formed through accretion-induced collapse may directly form MSPs \citep[e.g.,][]{Hurley:2010MNRAS,Tauris:2013A&A,Ruiter:2019MNRAS}, which are not the focus of this work. 
    Our models also do not include neutron stars formed from stellar merger products. These populations, whilst not expected to produce the majority of neutron stars, may go some way to explaining the low pulsar birth rates found in this study.

    \item Our model currently assumes a binary fraction of 100\% for massive stars. Whilst this is appropriate for massive O-type stars \citep[][]{Sana:2012Sci,Sana:2014ApJS}, B-type stars likely have somewhat lower binary fractions around 70\% \citep[][]{MoeDiStefano::2017ApJS,Zapartas:2017A&A}. Future work should incorporate a mass dependent binary fraction and the contribution of single stars.
    This will be particularly important hen considering millisecond pulsars and accretion induced collapses, as binary interactions play a more important role in the evolution of these systems. 

    \item Our model of an exponentially decaying magnetic field  (Section~\ref{subsec:pulsar_evolution}) is meant to provide a simple starting point for exploring the possibility of pulsar magnetic fields decaying on a broad range of timescales, as well as allowing for comparison to previous work \citep[][]{Kiel:2008MNRAS,Chattopadhyay2020,Dirson2022}. Recent work has introduced more sophisticated models for magnetic field decay motivated by detailed theoretical work \citep[e.g.,][]{Vigano:2021CoPhC,Sarin:2023ApJL,Graber:2024ApJ}. We leave the implementation of these models, and an exploration of their impact to future work.
    
    \item We currently lack a physical model for the birth spin periods and magnetic fields for pulsars as a function of the progenitor properties \citep[cf.][for neutron star masses and kicks]{Mandel:2020qwb}. 
    This work, and any follow-up in the foreseeable future will only be able to use some empirically assumed distribution of these quantities.

    \item We have neglected the population of magnetars with long spin periods and high magnetic fields \citep[e.g.,][]{Ferrario:2006MNRAS,Ferrario:2008MNRAS,Popov:2010MNRAS,Gullon:2015MNRAS,Makarenko:2021MNRAS}. As with canonical pulsars, most magnetars are isolated, whilst their massive star progenitors were presumably born in binaries. Binary interactions such as stellar mergers may be responsible for producing highly magnetic neutron stars \citep[][]{Ferrario:2009MNRAS,Schneider:2019Nature,Frost:2024Sci}.

    \item Our model assumes that the angle between the magnetic and spin axes of pulsars is constant for all pulsars and does not evolve with time. This is a simplification, and we will explore the implications of a varying angle in future work \citep[e.g.,][]{Johnston2020}.

    \item Updated treatment of both radio and \gray\ selection effects, including the dependence of pulsar radio luminosity and beam geometry on pulsar spin properties \citep[cf.][]{Posselt:2023MNRAS} will improve the fidelity of our models. 
\end{enumerate}

After addressing the above-mentioned caveats, we would be able to simulate full populations of NSs and broaden the scope of our research with several different follow-up studies: 
\begin{enumerate}
    \item A full scope of pulsar population synthesis can enable the study of NS contribution to the diffuse \gray\ background \citep{2017arXiv170500009F}.
    
    \item The inclusion of MSPs in the population synthesis models would provide further insight into the possible MSP origin of excess of GeV \gray\ emission from the Galactic Centre \citep{hooperm2016, ploeg2017, eckner2018, Gautam:2022NatAs}.
    
    \item A comparison of pulsar populations between the Galactic field and globular clusters with results from dynamical simulations, such as those from NBODY simulations \citep[e.g.,][]{Ye:2019ApJ,Ye:2024ApJ}. 
    
    \item After adding in the formation of MSPs and NSs formed through accretion-induced collapse, the model will be able to perform on the Hubble time scale. 
    This will enable the modelling of merger events for gravitational wave detections, such as black hole-NS binaries \citep[][]{Chattopadhyay2021}, double NSs \citep[][]{Chattopadhyay2020} and NS-white dwarf binaries. 
    
    \item Including stellar mergers and the physics of magnetar formation and evolution in future versions of COMPAS will allow us to synthesise magnetar populations and compare them to observations. This will open up avenues for comparing COMPAS simulations to observations of fast radio bursts.
    
    \item Investigate more sophisticated model comparison methods to fully explore the parameter space and identify a best-fit model that produces the population of pulsars observed to date.
    Several approaches have been taken in the literature, including hierarchical Bayesian parameter estimation \citep[e.g.,][]{Ploeg:2020JCAP} and machine learning approaches \citep[e.g.,][]{Ronchi:2021ApJ,Graber:2024ApJ,PardoAraujo:2025A&A}. At present, such a large parameter space exploration is computationally infeasible using COMPAS.
    In a follow-up study, we will optimise, parameterise and automate our COMPAS simulations and post-processing pipelines so that the use of more sophisticated model exploration and comparison techniques becomes computationally feasible.
\end{enumerate}

In future work, we aim to study how different binary properties, including initial orbital properties, different mass transfer theories, neutron star kick velocities and common envelope physics can impact the pulsar population, and how this population can be used to constrain the underlying physics of massive binary evolution.

\section{Conclusions}
\label{sec:conclusions}

We performed a population synthesis of the canonical Galactic pulsar population using the rapid binary population synthesis code COMPAS \citep[][]{COMPAS} on both radio and \gray\ canonical pulsars.
We generated 40 different models for pulsar birth properties for COMPAS runs incorporating a magnetic braking model with a braking index of 3 for pulsar spindown. 
We discover that the PN1-B1-D4 model, where the birth period of pulsar follows a normal distribution with $\mu = 75$ ms and $\sigma = 25$ ms, the birth magnetic field of pulsars follows a uniform distribution between $10^{11}$ and $10^{13}$\,G, and magnetic decay time scale $\tau_d = 1$\,Myr best describes the observed pulsar population, based on the facts that:

\begin{enumerate}

    \item PN1-B1-D4 model produces the number of radio and \gray\ canonical pulsars that, given the star formation rate of the model, best fits the catalogue. 

    \item The distribution of physical properties of radio and \gray\ pulsars produced from this model matches the catalogue the best. 
\end{enumerate}

We then compare in detail the PN1-B1-D4 model results to the catalogue by examining the P-\Pdot\ plots, the Galactocentric coordinates, and the heuristic flux between the simulation and the catalogue. 
We also discussed the impact of the selection effect on the population synthesis results, demonstrating that the slightest change in pulsar physics and observation sensitivity can bring significant differences to the final simulation. 

Using the results from PN1-B1-D4 model, we also attempted to explore the possible emission mechanism from low \Edot\ pulsars (\Edot\ $<10^{33}$ \LumUnit . Our results seem to favour the isotropic emission mechanism over the beamed emission mechanism. 

We will follow up this study to include the Galactic MSP population and utilise the results to answer other physical questions, including the \gray\ diffuse background, Galactic Centre GeV excess, and gravitational wave events from a NS binary.

\begin{acknowledgement}

The authors appreciate coding assistance from R. Wilcox and J. Riley, and helpful discussions from J. Hurley. 
\end{acknowledgement}

\paragraph{Funding Statement}

This research was conducted by the Australian Research Council (ARC) Centre of Excellence for Gravitational Wave Discovery (OzGrav), through project numbers CE170100004 and CE230100016. 
S. Stevenson is a recipient of an ARC Discovery Early Career Research Award (DE220100241). 
D. Chattopadhyay is supported by the UK's Science and Technology Facilities Council grant ST/V005618/1. 
This work used the OzSTAR and Ngarrgu Tindebeek (NT) high-performance computers at Swinburne University of Technology. OzSTAR and NT are funded by Swinburne University of Technology and the National Collaborative Research Infrastructure Strategy (NCRIS), and the Victorian Higher Education State Investment Fund (VHESIF).

\paragraph{Competing Interests}

None.

\paragraph{Data Availability Statement}

COMPAS is open source and hosted on Github (\url{https://github.com/TeamCOMPAS/COMPAS}). Due to the size of COMPAS output files, the authors can share configuration files used to produce the pop-synth results upon request. 

Our post-processing scripts are available on Github at \url{https://github.com/yuzhesong/compas\_pulsar\_paper.git}.

\printendnotes

\bibliography{main}

\end{document}